\documentclass[fleqn,usenatbib]{mnras}

\usepackage{newtxtext,newtxmath}

\usepackage[T1]{fontenc}

\DeclareRobustCommand{\VAN}[3]{#2}
\let\VANthebibliography\thebibliography
\def\thebibliography{\DeclareRobustCommand{\VAN}[3]{##3}\VANthebibliography}

\usepackage{subcaption} 
\usepackage{multirow} 

\usepackage{graphicx}	
\usepackage{amsmath}	

\newcommand{\kms}{\mathrm{km\,s}^{-1}} 
\newcommand{\density}{\mathrm{M_{\sun}\,pc^{-3}}}

\title[Binary black holes in dense clusters]{ Formation and evolution of binary black holes in $N$-body simulations of  star clusters with up to two million stars}

\author[J Barber et al.]{
Jordan Barber,$^{1}$\thanks{E-mail: barberj2@cardiff.ac.uk}
Fabio Antonini,$^{1}$
\\
$^{1}$Gravity Exploration Institute, School of Physics and Astronomy, Cardiff University, Cardiff , CF24 3AA, UK
}

\date{Accepted XXX. Received YYY; in original form ZZZ}

\pubyear{\the\year{}}

\begin{document}
\label{firstpage}
\pagerange{\pageref{firstpage}--\pageref{lastpage}}
\maketitle

\begin{abstract}
Understanding binary black hole (BBH) dynamics in dense star clusters is key to  interpreting the gravitational wave detections by LIGO and Virgo. Here, we perform $N$-body simulations of star clusters, focusing on BBH formation mechanisms, dynamical evolution and merging properties. We explore a wide parameter space of initial conditions, with cluster masses ranging from $10^{4}$ to $10^{6}~\mathrm{M_{\sun}}$, densities from $10^{3}$ to $10^{5}~\rm \density$, and up to $100\%$ of massive stars in binaries.
We show that most BBH mergers  originate from the primordial binary population rather than being dynamically assembled, and that the evolution towards merger for most of these binaries is not significantly altered by dynamical encounters.  As a result, the overall number of BBH mergers from the $N$-body simulations is nearly identical to that obtained when the same stellar population is evolved in isolation.
Contrary to theoretical expectations,  nearly all dynamically formed BBH mergers occur when the binary is still bound to its host cluster, with $\simeq 90\%$ of all dynamical mergers occurring within the cluster core region. In about half of these mergers the binary is part of a stable black hole-triple system.
In one  model,  stellar mergers lead to the formation of a $\simeq 200\,\mathrm{M_\odot}$ black hole, which then grows to $\simeq 300\,\mathrm{M_\odot}$ through black hole mergers. Our study highlights the importance of detailed $N$-body simulations in capturing the    evolution of black hole populations in dense clusters and challenges  conclusions  based on semi-analytical and Monte Carlo methods.
\end{abstract}




\section{Introduction}

The fourth observing run of the LIGO-Virgo-KAGRA collaboration is currently underway and thus far there have been $>120$ publicly announced confident detections, adding to the $90$ observed from the previous three observing runs \citep{abbott_observation_2016, abbott_binary_2016, abbott_gwtc-1_2019, abbott_gwtc-2_2021, the_ligo_scientific_collaboration_gwtc-3_2021}. Whilst the vast majority of these detections have been binary black hole  (BBH) mergers, the formation mechanism for these systems is still uncertain. 

Broadly speaking, there are two main BBH formation channels, an isolated evolution channel and a dynamical channel. Through the isolated channel, two massive stars are born in a bound binary and subsequently co-evolve in absence of strong external interactions, for example in the galactic field. During their evolution, this binary likely undergoes some period of common envelope evolution, which efficiently shrinks the binary separation \citep[e.g.,][]{tutukov_evolution_1973, dominik_double_2012, ivanova_common_2013, glanz_common_2021} Alternatively, the binary's orbit could tighten through stable mass transfer episodes, where matter flows from one star to its companion over extended periods \citep[e.g.,][]{van_den_heuvel_forming_2017, pavlovskii_stability_2017, neijssel_effect_2019, shao_stable_2022} Additionally, chemically homogeneous evolution, driven by efficient rotational mixing in rapidly spinning massive stars, can lead to the formation of close BBH systems without significant expansion of the stellar radii \citep[e.g.,][]{mandel_merging_2016, marchant_new_2016, dubuisson_cosmic_2020, riley_chemically_2021}. Eventually, each star will collapse into a black hole (BH), and provided the binary remains bound following the supernova kicks, the resulting BBH can  merge within a Hubble time. \citep[e.g.,][]{hurley_evolution_2002, belczynski_comprehensive_2002,de_mink_merger_2015, belczynski_effect_2016, spera_merging_2019, mapelli_binary_2020, belczynski_evolutionary_2020, broekgaarden_impact_2021, costa_formation_2021, qin_merging_2023}. 

The dynamical channel involves the formation and evolution of a BBH within a dense stellar system, where the binary experiences many gravitational encounters with other black holes (BHs) and stars. Through these interactions, BBHs can be formed and disrupted, and have their orbital properties altered through dynamical hardening. In certain cases, these encounters can bring a BBH into a regime where it is able to merge within a Hubble time. 

Since this mechanism requires dynamically active environments, the cores of dense stellar clusters such as globular clusters \citep[e.g.,][]{coleman_miller_production_2002, rodriguez_dynamical_2016, askar_mocca-survey_2017, samsing_eccentric_2018, hong_binary_2018, rodriguez_post-newtonian_2018, rodriguez_post-newtonian_2018-1, sedda_mocca-survey_2019,anagnostou_dynamically_2020, antonini_merger_2020, arca_sedda_merging_2021, leveque_mocca-survey_2023, torniamenti_hierarchical_2024, arcasedda_span_2024-1}, nuclear clusters \citep[e.g.,][]{miller_mergers_2009, antonini_secular_2012, antonini_merging_2016, bartos_rapid_2017,mapelli_hierarchical_2021, atallah_growing_2022, rodriguez_modeling_2022} and open clusters \cite[e.g.,][]{banerjee_stellar-mass_2018, dicarlo_merging_2019, di_carlo_binary_2020, dicarlo_intermediate-mass_2021, rastello_dynamics_2021, torniamenti_dynamics_2022, banerjee_binary_2022} are ideal locations. Nuclear clusters may also house an active galactic nucleus, where an accretion disc has formed around the central supermassive BH. In this case there are additional dynamical processes that can lead to BBH formation and merger within the AGN disc, such as torques exerted on the BBH by the surrounding dense gas, and binary accretion effects \citep[e.g.,][]{stone_assisted_2017, bartos_rapid_2017, mckernan_constraining_2018, grobner_binary_2020, fabj_eccentric_2024}.

The relative importance of the two formation channels in producing a detectable population of BBHs depends on two key factors: the proportion of stars that form in dense star clusters (and its redshift dependence), and  the efficiency of both stellar evolution and dynamical processes in driving BBHs to merge.
Most massive binaries are located in clusters and associations, which include both bound and unbound systems. However, the majority of these stars are not gravitationally bound to one another, and in many cases, the surrounding stellar densities are too low for dynamical interactions to play a significant role in their evolution \citep{krumholz_star_2019}. As a result, most massive starts can be considered as evolving in the field of the galaxy,  only influenced by their closest stellar companions -- the majority of black hole progenitors are found in binary or higher-multiplicity systems \citep[e.g.,][]{sana_binary_2012,moe_mind_2017}.
Conversely, in dense star clusters, dynamical processes can significantly enhance the merger rate of BBHs, making this channel a potentially important contributor to the observed BBH population. While there has been extensive theoretical work on how dynamical encounters might influence the number and properties of detected BBHs \citep{goodman_binarysingle-star_1993, miller_four-body_2002, samsing_black_2018-1} , reaching a clear conclusion remains challenging. This is mainly due to the complexity involved in simulating the relevant parameter space for star clusters in a self-consistent and accurate way.

Current star cluster simulations generally use one of three approaches: $N$-body simulations, Monte Carlo  methods, or Semi-analytical methods.
Direct $N$-body codes work by directly integrating the equations of motion for each star in the cluster, taking into account the gravitational interactions from all other stars. $N$-body simulations are the most accurate way to determine the evolution of a star cluster model and its BH population. However, the computational demands of this approach grow rapidly with the number of stars, making it feasible to simulate only relatively small systems, typically with $N \lesssim 10^5$ \citep[e.g.,][]{aarseth_1963, aarseth_1966, elena_1987, aarseth_1998, aarseth_2011, portegies_zwart_star_2001, wang_petar_2020}. 
In contrast, Monte Carlo codes use statistical techniques based on assumed models for the behaviour of BBHs and the overall cluster dynamics. This allows them to efficiently simulate much larger systems, with $N \gtrsim 10^5$, but at the cost of reduced accuracy. Monte Carlo methods struggle to resolve interactions that occur over timescales that are much shorter than  the cluster's relaxation time \citep{rodriguez_million-body_2016}. Additionally, they only account for interactions involving binaries and single stars. These interactions are typically calculated with high precision, using a direct integrator \citep{spurzem_1996, joshi2000monte, Giersz_2000, fregeau_monte_2007, giersz_2013, pattabiraman2013parallel, hypki_mocca_2013, rodriguez_post-newtonian_2018-1, hypki_2022, rodriguez_modeling_2022}.
Semi-analytical approaches, further approximate the calculation by using analytical formulae to describe how hard binaries evolve due to binary-single interactions, and how their production rate in the cluster is linked to the evolution of the cluster properties \citep{antonini_merging_2016,antonini_population_2020, antonini_merger_2020, mapelli_cosmic_2022,kritos_rapster_2022,arca_sedda_isolated_2023}.

Recent years have seen  several important improvements in the modelling of massive star clusters. $N$-Body simulations are starting to push the number of particles above the $N=10^{5}$ limit \citep{wang_nbody6gpu_2015, banerjee_binary_2022, arcasedda_span_2024}  with the first million particle N-body simulation performed in \cite{wang_2016_million}. A particular point of development moves $N$-body codes from complete particle-particle calculations to a hybrid particle-tree, particle-particle method which includes a regularisation scheme for the closest interactions \citep{iwasawa_gpu-enabled_2015,iwasawa_pentacle_2017, rantala_mstar_2020, rantala_frost_2021,rantala_bifrost_2023}. 
In this work, we use the highly efficient hybrid $N$-body code $\tt PeTar$ recently developed by \citet{wang_petar_2020}.
This code combines the particle-tree particle-particle method \citep{oshino_particleparticle_2011} and the slow-down algorithmic regularisation method \citep[SDAR][]{wang_slow-down_2020} to efficiently simulate the evolution of star clusters, while Stellar evolution is modelled using the single and binary stellar evolution packages \citep[$\tt SSE$ and $\tt BSE$ respectively;][]{hurley_comprehensive_2000, hurley_evolution_2002, banerjee_bse_2020}. This allows us to simulate clusters starting with masses and half-mass densities larger than has been previously explored and up to $10^6~\rm M_{\sun}$ and  $10^{5}~\rm \density$, respectively. This region of parameter space has thus far been sparsely sampled by previous work;  some work started at slightly higher densities but at much smaller cluster mass \citep{rastello_dynamics_2021, rizzuto_intermediate_2021, arca_sedda_2023, rantala_2024}. The wide range of initial conditions we consider in this work allows us to address two important questions: (i) what is the effect of dynamical encounters on the rate and properties of BBH mergers; and (ii) how these scale with the mass and density of the host cluster. More generally, we test our theoretical understanding of BBH formation in dynamical environments using  $N$-body simulations that  take into account both  gravitational interactions and stellar evolution processes and that contain a realistic initial population of stellar binaries.

The paper is organised as follows: Section~\ref{sec:methods} describes the initialisation of our clusters and methods; Section~\ref{sec:formation&merger} details the analysis of our results in terms of the overall population of BBHs and BHs within the clusters and the effect of dynamics on their properties.
In Section~\ref{sec:inclvseject} we compare the number of ejected and in-cluster mergers to those predicted by theoretical models.
In Section~\ref{sec:Massive} we investigate the formation of massive BHs in our simulations. Finally, Section~\ref{sec:conclusion} summarises the results from our study.

\begin{table*}
    \caption{Initial cluster conditions for our $\tt PeTar$ $N$-body simulations. Each model is given a unique name based on its initial setup (metallicity, initial cluster mass and density) with a -L added to models which are run for three $\rm Gyr$ instead of one $\rm Gyr$. Each model contains two variations, one which starts with no binaries, and one which sets an initial binary fraction of 100\% amongst massive stars (initial mass $\geq20~\rm M_{\sun}$). Finally the * denotes a model with a \citet{duquennoy_multiplicity_1991} period distribution instead of the \citet{sana_binary_2012} distribution used in all the other models.}
    \label{tab:initCond}
    \centering
    \begin{tabular}{ccccccccc}
        Model & Metallicity & Total Mass & Density & Half-Mass Relaxation Time & Binary Fraction & End Time & Binary Period Dist \\
         & & $\rm M_{\sun}$ & $\rm \density$ & $\rm Myr$ & & $\rm Myr$ & \\ \hline
        \multirow{2}{*}{Z1-M1-D3} & \multirow{6}{*}{0.01} & \multirow{2}{*}{10,000} & \multirow{2}{*}{1200} & \multirow{2}{*}{11.5} & 0 & \multirow{2}{*}{1000} & \multirow{2}{*}{Sana}\\
         & & & & & 0.0025 & & \\
        \multirow{2}{*}{Z1-M5-D3} & & \multirow{2}{*}{50,000} & \multirow{2}{*}{1200} & \multirow{2}{*}{47.0} & 0 & \multirow{2}{*}{1000} & \multirow{2}{*}{Sana} \\
         & & & & & 0.0025 &  &  \\
        \multirow{2}{*}{Z1-M10-D3} & & \multirow{2}{*}{100,000} & \multirow{2}{*}{1200} & \multirow{2}{*}{86.2} & 0 & \multirow{2}{*}{1000} & \multirow{2}{*}{Sana} \\
         & & & & & 0.0026 & &  \\
         Z1-M50-D3 & & 500,000 & 1200 & 253.7 & 0 & 608 & Sana \\
         Z1-M100-D3 & & 1,000,000 & 1200 & 506.8 & 0 & 632 & Sana \\ \hline
        \multirow{2}{*}{Z2-M1-D3} & \multirow{16}{*}{0.001} & \multirow{2}{*}{10,000} & \multirow{2}{*}{1200} & \multirow{2}{*}{11.3} & 0 & \multirow{2}{*}{1000} & \multirow{2}{*}{Sana} \\
         & & & & & 0.0025 & &  \\
        \multirow{2}{*}{Z2-M5-D3} & & \multirow{2}{*}{50,000} & \multirow{2}{*}{1200} & \multirow{2}{*}{49.4} & 0 & \multirow{2}{*}{1000} & \multirow{2}{*}{Sana} \\
         & & & & & 0.0025 & &  \\
        \multirow{2}{*}{Z2-M5-D3-L} & & \multirow{2}{*}{50,000} & \multirow{2}{*}{1200} & \multirow{2}{*}{49.4} & 0 & \multirow{2}{*}{3000} & \multirow{2}{*}{Sana} \\
         & & & & & 0.0025 & &  \\
        \multirow{2}{*}{Z2-M10-D3} & & \multirow{2}{*}{100,000} & \multirow{2}{*}{1200} & \multirow{2}{*}{86.2} & 0 & \multirow{2}{*}{1000} & \multirow{2}{*}{Sana} \\
         & & & & & 0.0026 & & \\
        \multirow{2}{*}{Z2-M10-D3-L} & & \multirow{2}{*}{100,000} & \multirow{2}{*}{1200} & \multirow{2}{*}{86.2} & 0 & \multirow{2}{*}{3000} & \multirow{2}{*}{Sana} \\
         & & & & & 0.0026 & &  \\
        \multirow{2}{*}{Z2-M10-D3-L*} & & \multirow{2}{*}{100,000} & \multirow{2}{*}{1200} & \multirow{2}{*}{86.2} & 0 & \multirow{2}{*}{3000} & \multirow{2}{*}{Duquennoy \& Mayor} \\
         & & & & & 0.0026 & &  \\ 

        \multirow{2}{*}{Z2-M10-D4} & & \multirow{2}{*}{100,000} & \multirow{2}{*}{10,000} & \multirow{2}{*}{24.4} & 0 & \multirow{2}{*}{1000} & \multirow{2}{*}{Sana} \\
         & & & & & 0.0025 & &  \\
        
        \multirow{2}{*}{Z2-M1-D5} & & \multirow{2}{*}{10,000} & \multirow{2}{*}{100,000} & \multirow{2}{*}{0.561} & 0 & \multirow{2}{*}{1000} & \multirow{2}{*}{Sana} \\
         & & & & & 0.0025 & &  \\
        \multirow{2}{*}{Z2-M5-D5} & & \multirow{2}{*}{50,000} & \multirow{2}{*}{100,000} & \multirow{2}{*}{2.78} & 0 & \multirow{2}{*}{1000} & \multirow{2}{*}{Sana} \\
         & & & & & 0.0025 & & \\\hline
        \multirow{2}{*}{Z3-M1-D3} & \multirow{6}{*}{0.0001} & \multirow{2}{*}{10,000} & \multirow{2}{*}{1200} &   \multirow{2}{*}{11.2} & 0 & \multirow{2}{*}{1000} & \multirow{2}{*}{Sana} \\
         & & & & & 0.0025 & &  \\
        \multirow{2}{*}{Z3-M5-D3} & & \multirow{2}{*}{50,000} & \multirow{2}{*}{1200} & \multirow{2}{*}{47.7} & 0 & \multirow{2}{*}{1000} & \multirow{2}{*}{Sana} \\
         & & & & & 0.0025 & &  \\
        \multirow{2}{*}{Z3-M10-D3} & & \multirow{2}{*}{100,000} & \multirow{2}{*}{1200} & \multirow{2}{*}{86.7} & 0 & \multirow{2}{*}{1000} & \multirow{2}{*}{Sana} \\
         & & & & & 0.0025 & &  \\ 
         Z3-M50-D3 & & 500,000 & 1200 & 253.7 & 0 & 568 & Sana \\
         Z3-M100-D3 & & 1,000,000 & 1200 & 506.8 & 0 & 280 & Sana \\
         \hline
    \end{tabular}
\end{table*}

\begin{table*}
    \caption{Here we show the end state information from our simulations. The first two columns denote the cluster model name (defined in Table~\ref{tab:initCond}) and whether this cluster is initialised with a primordial binary fraction. Columns 3 and 4 denote the number of BHs and BBHs within the cluster at the end of the simulation time, with the BBH column further split into the total number of BBHs, the number of hard BBHs and the number of binary systems with only one BH. Columns 5, 6 and 7 describe the merging BBHs coming from the primordial binaries, dynamically formed binaries and the combined total respectively. Each of these groups is further split into the total number for that group, the number of in-cluster mergers and the number of ejected mergers. The final column shows the total merger efficiency for each cluster.}
    \label{tab:finalstate}
    \begin{tabular}{cccccccc}
        Model & \multicolumn{1}{c|}{} & $N_{\mathrm{sing}}$ & \begin{tabular}[c]{@{}c@{}}$N_{\mathrm{BHBs}}$\\ Tot(Hard)Star\end{tabular} & \begin{tabular}[c]{@{}c@{}}Primordial\\ Mergers \\ Tot(Incl)Ejec\end{tabular} & \begin{tabular}[c]{@{}c@{}}Dynamical\\ Mergers \\ Tot(Incl)Ejec\end{tabular} & \begin{tabular}[c]{@{}c@{}}Total\\ Mergers \\ Tot(Incl)Ejec\end{tabular} & \begin{tabular}[c]{@{}c@{}}Merger\\ Efficiency\end{tabular} \\ \hline
        \multirow{2}{*}{Z1-M1-D3} & \multicolumn{1}{c|}{With Binaries} & 4 & 1(1)0 & 2(0)2 & 0(0)0 & 2(0)2 & $2.0\times10^{-4}$ \\
         & No Binaries & 0 & 1(1)0 & - & 2(2)0 & 2(2)0 & $2.0\times10^{-4}$ \\
        \multirow{2}{*}{Z1-M5-D3} & With Binaries & 23 & 1(1)2 & 4(0)4 & 0(0)0 & 4(0)4 & $8.0\times10^{-5}$ \\
         & No Binaries & 28 & 1(1)1 & - & 3(2)1 & 3(2)1 & $6.0\times10^{-5}$ \\
        \multirow{2}{*}{Z1-M10-D3} & With Binaries & 50 & 1(1)7 & 14(6)8 & 4(3)1 & 18(9)9 & $1.8\times10^{-4}$ \\
         & No Binaries & 95 & 2(1)2 & - & 7(6)1 & 7(6)1 & $7.0\times10^{-5}$ \\ 
        Z1-M50-D3 & No Binaries & 404 & 1(1)4 & - & 3(3)0 & 3(3)0 & $6.0\times10^{-6}$\\
         Z1-M100-D3 & No Binaries & 780 & 0(0)3 & - & 0(0)0 & 0(0)0 & - \\\hline
        
        \multirow{2}{*}{Z2-M1-D3} & With Binaries & 3 & 3(3)1 & 2(1)1 & 1(0)1 & 3(1)2 & $3.0\times10^{-4}$ \\
         & No Binaries & 3 & 1(1)0 & - & 1(1)0 & 1(1)0 & $1.0\times10^{-4}$ \\
        \multirow{2}{*}{Z2-M5-D3} & With Binaries & 29 & 1(1)3 & 11(5)6 & 4(4)0 & 15(9)6 & $3.0\times10^{-4}$ \\
         & No Binaries & 50 & 3(2)0 & - & 1(1)0 & 1(1)0 & $2.0\times10^{-5}$ \\
        \multirow{2}{*}{Z2-M10-D3} & With Binaries & 70 & 1(1)4 & 30(10)20 & 5(4)1 & 35(14)21 & $3.5\times10^{-4}$ \\
         & No Binaries & 108 & 2(1)2 & - & 8(8)0 & 8(8)0 & $8.0\times10^{-5}$ \\
        \multirow{2}{*}{Z2-M5-D3-L} & With Binaries & 5 & 2(2)4 & 8(5)3 & 6(6)0 & 14(11)3 & $2.8\times10^{-4}$ \\
         & No Binaries & 28 & 2(2)0 & - & 4(3)1 & 4(3)1 & $8.0\times10^{-5}$ \\
        \multirow{2}{*}{Z2-M10-D3-L} & With Binaries & 32 & 4(3)2 & 5(1)4 & 4(4)0 & 9(5)4 & $9.0\times10^{-5}$ \\
         & No Binaries & 72 & 2(2)3 & - & 3(3)0 & 3(3)0 & $3.0\times10^{-5}$ \\
        \multirow{2}{*}{Z2-M10-D3-L*} & With Binaries & 84 & 10(9)3 & 12(4)8 & 4(3)1 & 16(7)9 & $1.6\times10^{-4}$ \\
         & No Binaries & 157 & 1(0)0 & - & 4(4)0 & 4(4)0 & $4.0\times10^{-5}$ \\
        
        \multirow{2}{*}{Z2-M10-D4} & With Binaries & 29 & 2(2)3 & 27(12)15 & 9(8)1 & 36(20)16 & $3.0\times10^{-4}$ \\
        & No Binaries & 82 & 1(1)2 & - & 10(9)1 & 10(9)1 & $1.0\times10^{-4}$ \\
        
        \multirow{2}{*}{Z2-M1-D5} & With Binaries & 0 & 0(0)1 & 3(0)3 & 2(2)0 & 5(2)3 & $5.0\times10^{-4}$ \\
         & No Binaries & 0 & 1(1)0 & - & 2(2)0 & 2(2)0 & $2.0\times10^{-4}$ \\
        \multirow{2}{*}{Z2-M5-D5} & With Binaries & 2 & 0(0)0 & 16(7)9 & 4(4)0 & 20(11)9 & $4.0\times10^{-4}$ \\
         & No Binaries & 6 & 1(1)3 & - & 4(3)1 & 4(3)1 & $6.0\times10^{-5}$ \\ \hline
        
        \multirow{2}{*}{Z3-M1-D3} & With Binaries & 2 & 0(0)1 & 6(3)3 & 1(1)0 & 7(4)3 & $7.0\times10^{-4}$ \\
         & No Binaries & 0 & 1(1)0 & - & 1(1)0 & 1(1)0 & $1.0\times10^{-4}$ \\
        \multirow{2}{*}{Z3-M5-D3} & With Binaries & 34 & 2(2)1 & 18(9)9 & 4(3)1 & 22(12)10 & $4.4\times10^{-4}$ \\
         & No Binaries & 50 & 2(2)0 & - & 3(3)0 & 3(3)0 & $6.0\times10^{-5}$ \\
        \multirow{2}{*}{Z3-M10-D3} & With Binaries & 95 & 6(5)4 & 28(13)15 & 6(5)1 & 34(18)16 & $3.4\times10^{-4}$ \\
         & No Binaries & 137 & 2(2)0 & - & 7(7)0 & 7(7)0 & $7.0\times10^{-5}$ \\ 
         Z3-M50-D3 & No Binaries & 777 & 3(1)5 & - & 6(5)1 & 6(5)1 & $1.2\times10^{-5}$ \\
         Z3-M100-D3 & No Binaries & 1641 & 0(0)0 & - & 0(0)0 & 0(0)0 & - \\\hline
    \end{tabular}
\end{table*}

\section{Methods} \label{sec:methods}
To study the formation mechanisms of BBHs we run 35 $N$-body cluster models. We use the high-performance hybrid $N$-body code $\tt PeTar$ \citep{wang_petar_2020}, which combines the particle-tree particle-particle method \citep[$\mathrm{P^{3}T}$; ][]{oshino_particleparticle_2011} and the slow-down algorithmic regularisation method \citep[$\tt SDAR$][]{wang_slow-down_2020} with parallisation via a hybrid parallel method based on the $\tt FDPS$ framework \citep{iwasawa_implementation_2016, iwasawa_accelerated_2020, namekata_fortran_2018}. In addition to OpenMP and MPI processes for parallisation, we chose 
a $\tt PeTar$ configuration which accelerates the long-range force calculation with Nvidea P100 GPUs. $\tt PeTar$ is computationally much more efficient than standard direct $N$-body codes \citep{wang_petar_2020}, allowing us to  explore a broader parameter space of initial conditions than previous work and to include high binary fractions approaching 100\% for massive stars. 

A drawback of $\tt PeTar$ is that it does not directly include post-Newtonian (PN) terms in the equations of motion, unlike some other $N$-body codes \citep{aarseth_mergers_2012}. Mergers through GW radiation are modelled  by computing the semi-major axis and eccentricity evolution described as in \citet{peters_gravitational_1964}:

\begin{equation}
    t_{\rm delay}(a_\mathrm{0}, e_\mathrm{0}) = \frac{12}{19}\frac{c^{4}_{0}}{\beta}\times\int^{e_{0}}_{0}de\frac{e^{29/19}[1+(121/304)e^{2}]^{1181/2299}}{(1-e^{2})^{3/2}},
    \label{eq:GWTime}
\end{equation}
where,
\begin{equation}
    c_{0} = \frac{a_{0}(1-e_{0}^{2})}{e_{0}^{12/19}}\left[1+\frac{121e_{0}^{2}}{304}\right]^{-870/2299}
\end{equation}
and,
\begin{equation}
    \beta=\frac{64}{5}\frac{G^{3}m_{1}m_{2}(m_{1}+m_{2})}{c^{5}}.
\end{equation}
To determine whether a binary merges, $\tt PeTar$ compares the GW timescale to the binary integration time-step. This time-step depends on strength of the perturbation from the binaries neighbours and the slowdown factor for the binary. The stronger the perturbation, the smaller this time-step becomes. When the GW timescale becomes shorter than the integration time-step, the binary is considered to have merged before the next time-step. Once this criteria is satisfied, the binary position is  evolved in space until the time of merger, and as such we can  find the actual position of a BBH merger in our clusters. 

Within the simulation we include stellar evolution of the stars using the single and binary stellar evolution packages \citep[$\tt SSE$ and $\tt BSE$ respectively;][]{hurley_comprehensive_2000, hurley_evolution_2002, banerjee_bse_2020}. Within BSE, compact objects can merge through gravitational wave emission. This is accounted for by computing the semi-major axis and eccentricity evolution using equation~(\ref{eq:GWTime}).

We consider a binary as \textit{ejected} from the cluster if the following  two conditions are satisfied. Firstly, we impose a distance criterion $r_{\mathrm{COM}}>20r_{\mathrm{h}}$, where $r_{\mathrm{COM}}$ is the centre of mass  position of the binary and $r_{\mathrm{h}}$ is the cluster half-mass radius at any given time. Secondly, there is an energy criterion $K_{\mathrm{COM}} + \Omega_{\mathrm{COM}} > 0$, where $K_{\mathrm{COM}}$ and $\Omega_{\mathrm{COM}}$ are the kinetic energy and potential energy of the binary centre of mass, respectively. 

In this work we use $\tt PeTar$ to explore astrophysically motivated initial cluster conditions in areas of the parameter space where there has been a lack of previous simulations.

\begin{figure*}
    \centering
    \includegraphics[width=2\columnwidth]{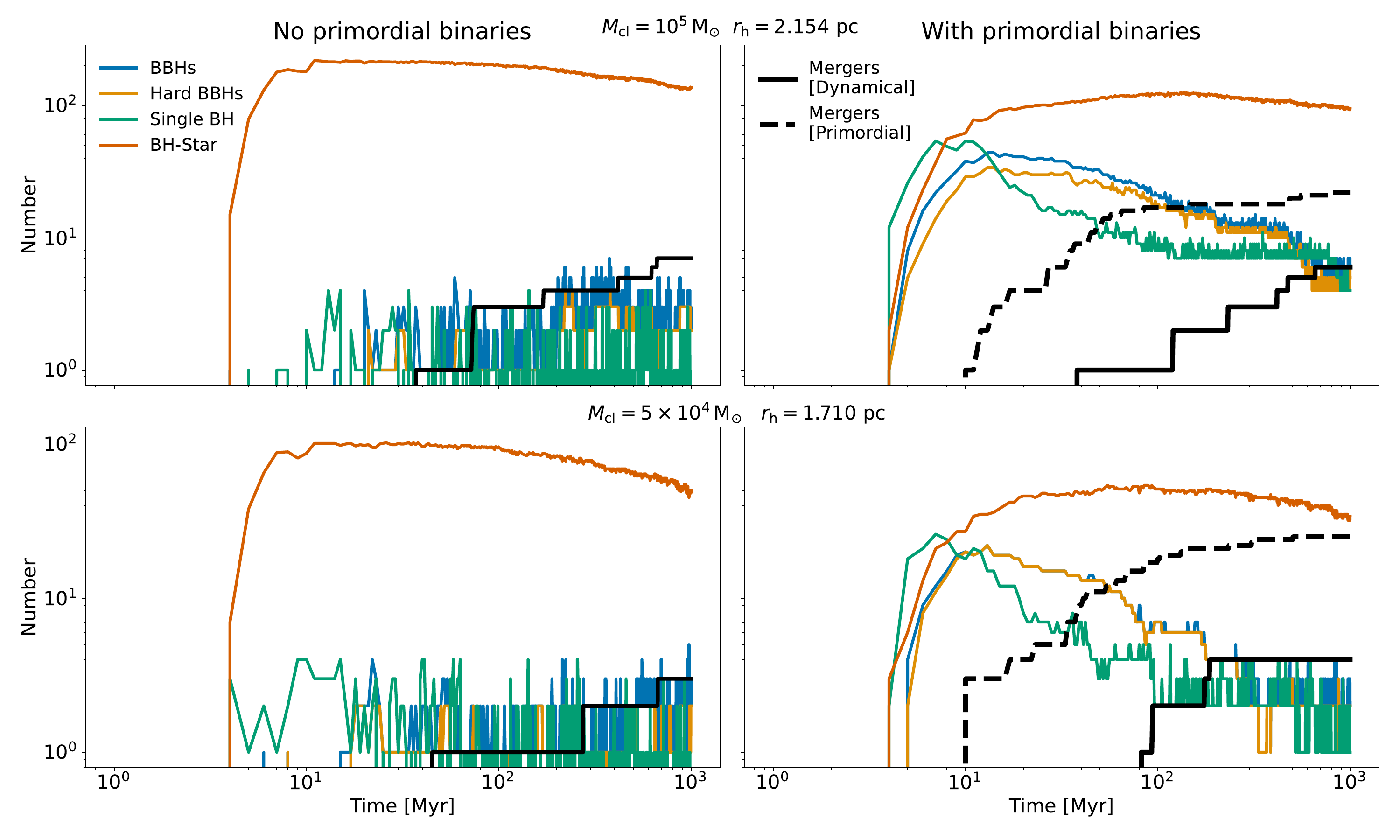}
    \caption{Number of BH and BBH population inside the cluster as a function of time. The BBH population is split into all BBHs, hard BBHs and binary systems containing only one BH. We over-plot the cumulative count of merging BBHs from the dynamically formed BBHs (solid line) and BBHs from the primordial binary population (dashed line). The upper panel shows Model Z3-M10-D3 while the bottom panel shows model Z3-M5-D3. We show both variations with(right) and without(left) a primordial binary population.}
    \label{fig:BHNumbers}
\end{figure*}

\begin{figure*}
    \centering
    \includegraphics[width=2\columnwidth]{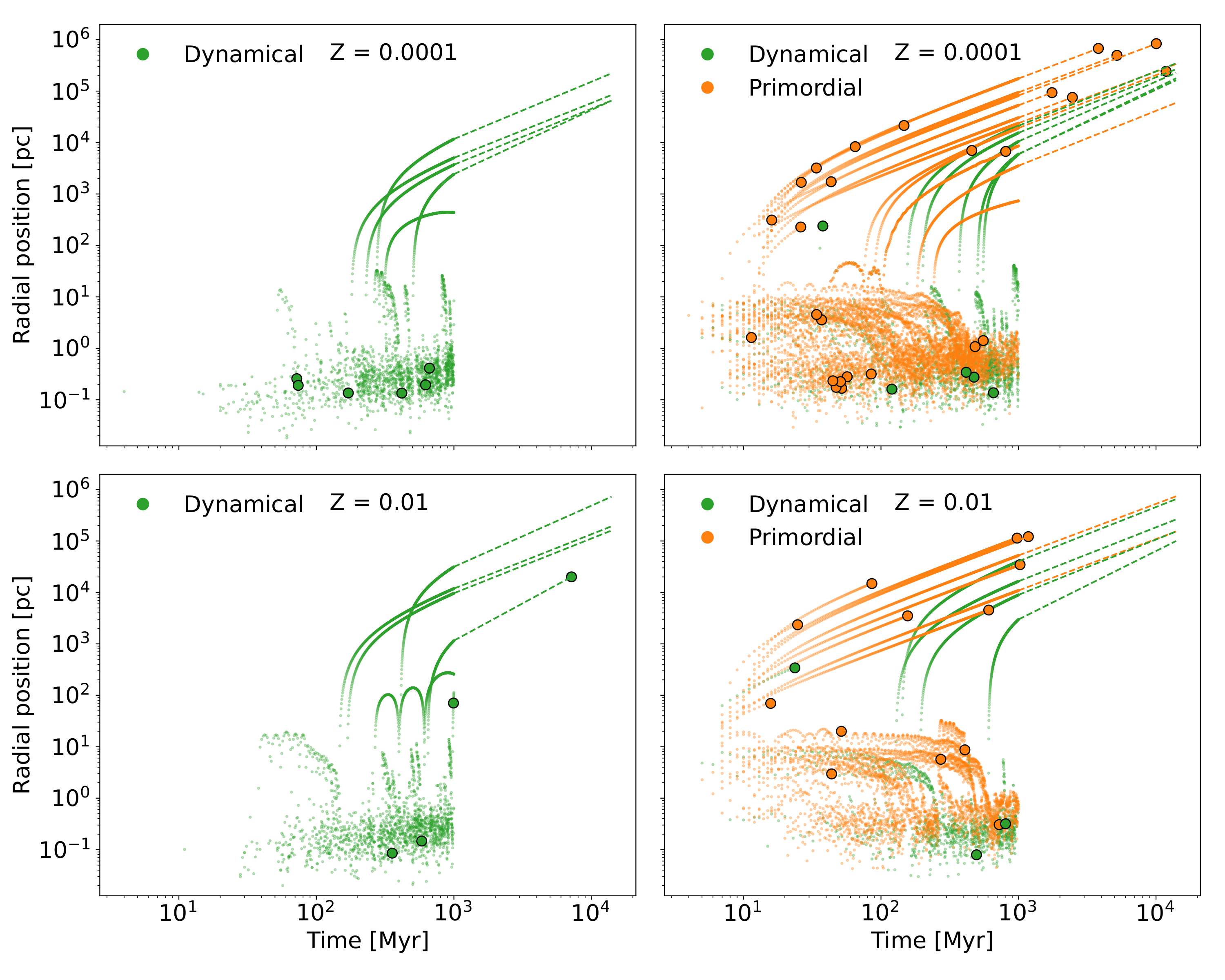}
    \caption{The evolution of the radial position of every BBH in the simulation. The top panels show  model Z3-M10-D3 and the bottom panels show model Z1-M10-D3, whilst the columns, right/left, show the cluster variation with/without a primordial binary population. Here we distinguish between the dynamically formed BBHs and those from the primordial binary population and highlight the time when the BBH mergers occur (filled circles). For the ejected population, we compute their delay time and plot the future path up to merger or up to $t_{\rm H}$.}
    \label{fig:RadvsTime}
\end{figure*}

\subsection{Initial Conditions}

We generate the cluster initial conditions using $\tt McLuster$ \citep{wang_complete_2019,kupper_mass_2011}. For every cluster we adopt a King density profile \citep{king_structure_1966} with a concentration parameter $W_{0}=8$ and assume that the cluster is not in any larger galactic tidal field. This value of $W_{0}$ allows us to explore the dynamics within compact clusters. Although not the focus of this study, we note that the choice of this parameter can significantly influence the formation of an intermediate mass BH (IMBH) \citep{rizzuto_intermediate_2021}. Most models have an initial half-mass density $\rho_{\mathrm{h}}=1.2\times10^{3}\,\density$ as this is a typical value found for globular clusters in the Galaxy \citep{harris_catalog_2013}.
We also explore higher densities, $\rho_{\mathrm{h}}=10^{4}\,\density$ and $\rho_{\mathrm{h}}=10^{5}\,\density$, since clusters might have been much denser in the past, and vary the initial cluster mass  from $10^{4}\,\mathrm{M_{\sun}}$ to $10^{6}\,\mathrm{M_{\sun}}$. We consider three values of metallicity: $Z=0.01,\,0.001\,\mathrm{and}\,0.0001$. 

To characterize the impact of an initial binary population on cluster evolution and BBH mergers, we consider clusters where all BH stellar progenitors start in a binary, which we refer to as primordial binaries.
For most cluster models we consider two variations. One variation begins with no primordial binaries, whilst in the other we ensure that every star with initial mass $>20\,\mathrm{M_{\sun}}$ is initialised in a binary. We opted against initialising lower mass binaries for three main reasons. Firstly, this work is focused on the effect of primordial binaries  on the overall rate of BBH mergers and on the formation and evolution of BBHs, and these are unlikely to be affected by binaries with low mass components. Secondly, it has been suggested that once BHs have been formed in the cluster, the dynamical evolution of the cluster properties depends only on the number and properties of the massive binaries \citep{wang_impact_2021}. Finally, the exclusion of  lower mass binaries makes our simulations  computational more efficient, reducing significantly the computing cost.

We sample the initial  masses of the cluster stars from a \citet{kroupa_variation_2001} initial mass function between $M = 0.08~\rm M_{\sun}$ and $150~\rm M_{\sun}$. Primordial binaries are then generated by taking every stellar mass $>20\,\mathrm{M_{\sun}}$ and drawing from a uniform mass ratio ($q$) distribution $0.1\leq q \leq1$; the particle in the cluster with the closest mass to what was drawn is then chosen as the binary partner. The eccentricity for these binaries is then drawn from a \citet{sana_binary_2012} distribution
\begin{equation}
    f_{\mathrm{e}}=0.55e^{-0.45}.
    \label{eq:eccdistinit}
\end{equation}
The binary period is set using the extended \citet{sana_binary_2012} distribution described in \citet{oh_dependency_2015}
\begin{equation}
    f_{\log_{10}(P)} = 0.23 \left[\log_{10}\left(\frac{P}{\rm days}\right)\right]^{-0.55}.
    \label{eq:periodinit}
\end{equation}
We also consider a single cluster variation where we  set the binary period based on the \citet{duquennoy_multiplicity_1991} distribution. This allows us to determine whether our results are particularly sensitive to the choice of period distribution.

We draw the supernova (SN) natal kicks from a Maxwellian distribution with $\sigma=265\,\kms$ \citep{hobbs_statistical_2005} and assume a fallback kick prescription when scaling the kicks for BH formation \citep{fryer_mass_1999}. In addition, we assume the rapid SN mechanism \citep{fryer_compact_2012} for compact object formation, and strong pulsational pair instability (PPSN) cut-off at $45\,\mathrm{M_{\odot}}$ \citep{belczynski_evolutionary_2020}.

In Table~\ref{tab:initCond} we summarise the initial conditions of all of our simulations. We also give the initial half-mass relaxation time for the cluster and the total physical time over which they are simulated. The half-mass relaxation time is given by \citep{spitzer_random_1971}

\begin{equation}
    t_{\mathrm{rh}} = \frac{0.138}{\langle m_{\mathrm{all}}\rangle\psi\log\Lambda} \sqrt{\frac{M_{\mathrm{cl}}r^{3}_{\mathrm{h}}}{G}},
    \label{eq:relax}
\end{equation}
where $M_{\mathrm{cl}}$ and $r_{\mathrm{h}}$ are the initial cluster mass and half-mass radius respectively, $\langle m_{\mathrm{all}}\rangle$ is the average stellar mass within $r_{\mathrm{h}}$ computed from the starting conditions of each cluster simulation, $\log\Lambda$ is the Coulomb logarithm, and $\psi$ depends on the mass spectrum within the half mass radius which we set 
equal to $5$ \citep[e.g.,][]{antonini_black_2019}.
The final integration time of the simulations is chosen such that it is several times the initial relaxation time of the cluster. For the most massive clusters ($M_{\rm cl}=5\times 10^5~\rm M_{\sun}$ and $10^6~\rm M_{\sun}$), we make sure that the simulation runs for at least one initial relaxation time. This means that all  cluster models have undergone significant dynamical evolution due to two-body relaxation by the end of the simulation. 

We choose some naming scheme for our simulations based on the metallicity, mass and density of the initial cluster for ease of reference. From here on we refer to specific simulations based on this identifier and whether the model contains primordial binaries. 

\begin{figure}
    \centering
    \includegraphics[width=0.5\textwidth]{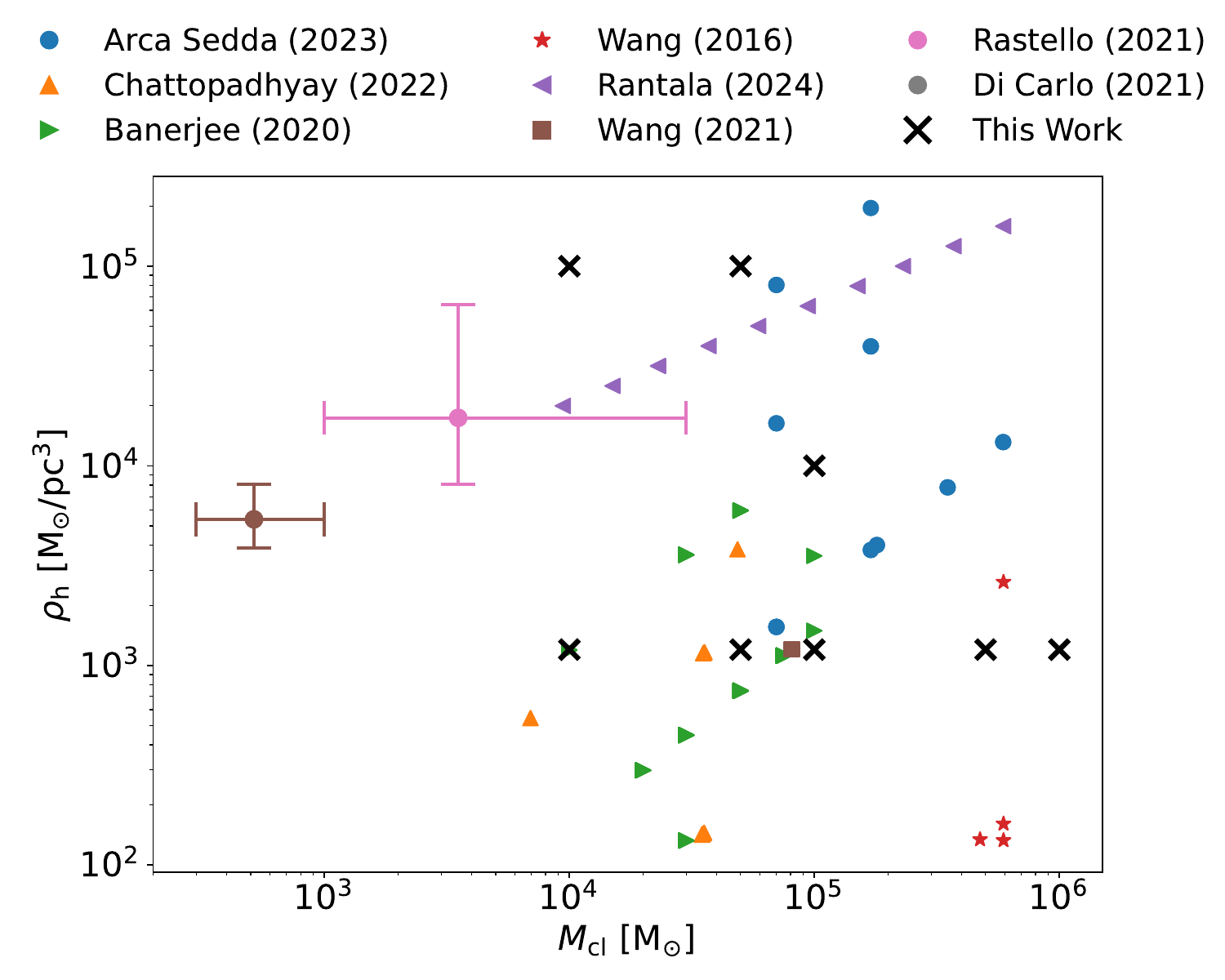}
    \caption{Comparison of the initial cluster properties ($\rho_{\rm h}$ and $M_{\rm cl}$) between the simulations in our work (black crosses) and previous studies; \citet{arca2024dragon} (blue dots), \citet{chattopadhyay_dynamical_2022} (orange dots), \citet{banerjee_stellar-mass_2020} (green dots), \citet{rastello_dynamics_2021} (red error bars) and \citet{dicarlo_intermediate-mass_2021} (pink error bars).}
    \label{fig:initialCond}
\end{figure}

In Fig.~\ref{fig:initialCond} we present the initial cluster conditions ($\rho_{\rm h}$ and $M_{\rm cl}$) simulated in this study, compared to those produced in previous $N$-Body studies\footnote{Where $\rho_{\rm h}$ is not directly given in these works, we compute it from given $M_{\rm cl}$ and $r_{\rm h}$ using the standard formula $\rho_{\rm h}=(M_{\rm cl}/2)/(4/3 \cdot\pi r_{\rm h}^{3})~.$} \citet{wang_2016_million,banerjee_stellar-mass_2020, rastello_dynamics_2021, wang_impact_2021, dicarlo_intermediate-mass_2021, chattopadhyay_dynamical_2022, arca_sedda_2023, rantala_2024}. This plot highlights the growing effort to numerically simulate clusters with high initial half-mass density and cluster mass. Our work then builds on these works to further populate this region of the parameter space with new models.

\section{Binary black hole Formation and Merger}\label{sec:formation&merger}

Through stellar evolution, the most massive stars become BHs in within $\lesssim 5\,\mathrm{Myr}$ whilst the smaller stars ($\sim20~\rm M_{\sun}$) collapse on a time scale $t\sim15~\rm Myr$, especially within metal rich environments. Once the BHs form inside the cluster they will either contribute to the single BH population or the BBH population\footnote{For now we ignore higher multiplicity systems i.e., triples, quadruples etc.}. BBHs are then either characterised as \textit{hard} BBHs or \textit{soft} BBHs depending on their binding energy compared to the average kinetic energy of stars in their immediate environment \citep{heggie_binary_1975}. We define the hard/soft boundary as:
\begin{equation}
    a_{\mathrm{h}} = \frac{GM_{\mathrm{tot}}}{\sigma^{2}}~,
    \label{eq:ah}
\end{equation}
 where $M_{\mathrm{tot}}$ is the total binary mass and $\sigma$ is the 1D velocity dispersion of the surrounding stellar objects, which is computed for each defined Lagrangian radii by $\tt PeTar$ in a data post-processing step. A binary with a separation $a<a_{\mathrm{h}}$ is considered a hard binary. 

After all massive stars have collapsed to BHs, the number of single BHs and BBHs  within a cluster will still evolve with time due to dynamical processes. New BBHs can be formed or disrupted through dynamical encounters, whilst existing BBHs may merge into a more massive BH. Dynamical encounters can also be responsible for ejecting both BHs and BBHs from a cluster \citep{morscher_dynamical_2015}. In Fig.~\ref{fig:BHNumbers} we show the evolution of these two populations for cluster models Z3-M10-D3 (upper panels) and Z3-M5-D3 (lower panels), we then show cluster variations with and without primordial binaries on the left and right panels, respectively. In these plots we also show the subset of hard BBHs as well as the number of binaries containing only one BH (which we refer to as BH-Star binaries). 

As expected, in all models we see the formation of the first BHs at $\approx4\,\mathrm{Myr}$ with a significant number of BBHs and BH-Star also forming at this time in models with a primordial binary population. In models without a primordial binary population,  BBHs are formed through dynamical interactions in the cluster core. Three-body binary formation processes  lead to the formation of the first BBHs in the cluster, which occurs approximately after a core-collapse time \citep{lee_evolution_1995}. 
After the first binary is formed, the number of BBHs within the cluster remains of $\mathcal{O}(1)$. A classical explanation for this is that once a BBH forms in the cluster core, it dominates the interactions, restricting further BBH formation and becoming a major energy source to the cluster \citep{heggie_gravitational_2003}. However, recent work has suggested an alternative explanation. When considering a high rate of binary-binary interactions in the core which efficiently ionise one of the binaries involved in the encounter \citep{marin_pina_dynamical_2023} the long term formation of multiple binaries is limited.

Clusters with a primordial binary population (right panels of Fig.~\ref{fig:BHNumbers}) form BBHs   much earlier on. This is because primordial binaries with sufficiently massive components and that remain bound through stellar evolution,  become BBHs after a time between $4\,\mathrm{Myr}$ to $\approx 10\,\mathrm{Myr}$. At this time
and in both cluster models, there are approximately as many BHs in BBHs as single BHs. It is expected that when a cluster contains a large number of BBHs, interactions between binaries {\rm or even higher multiplicity systems} can become the dominant form of encounters \citep{barber_black_2023}. These encounters are often chaotic with numerous potential end-states \citep{zevin_eccentric_2019}, including being a mechanism for stable triple BH formation \citep{sigurdsson_binarysingle_1993}, as discussed further in Section~\ref{sec:Dynms}. But importantly, they often lead to the disruption or ejection of binaries, thus reducing the number of BBHs  in the cluster. This is what we see in our models, to the extent that at the end of the simulation the number of BBHs  is largely independent of whether the cluster originally had a primordial binary population.
Table~\ref{tab:finalstate} summarises the end state for every cluster simulation we run, giving the final number of single BHs, BH-Star binaries and BBHs.

Fig.~\ref{fig:BHNumbers} also shows that in each simulation the hard BBHs represent a significant subset of the overall BBH population. For clusters with a primordial binary population, this is a result  of both stellar and dynamical processes. Firstly,   primordial stellar binaries that form with relatively close separations are likely to undergo further stellar interactions, causing their orbits to shrink. This process ultimately leads to the formation of BBHs with  high binding energies. Secondly, 
these interactions  also disrupt the soft BBHs over time, which explains why the relative fraction of hard BBHs increases with time and why near the end of the simulations  almost all BBHs in the cluster are hard (see Table~\ref{tab:finalstate}). 
For clusters without primordial binaries, BBHs can only form through dynamical encounters involving more than two BHs. Fig.~\ref{fig:BHNumbers} and Table~\ref{tab:finalstate}  show the evolution and final counts of BH-Star binaries, which whilst are not the focus of this work can be of interest with the recent \citet{gaia_collaboration_discovery_2024} BH observations.

In Fig.~\ref{fig:BHNumbers} we also display  the cumulative number of BBH mergers produced from each cluster simulation. 
We distinguish between two populations to which we will often refer to through the reset of the article: (1) BBH mergers in which the black hole components formed from stars that were originally part of the same binary system; and (2) BBHs that  paired through dynamical encounters. This distinction helps illustrate the different pathways through which BBH mergers can occur within the star cluster models.
Interestingly,  we see that the presence of primordial binaries has a negligible effect on the total number of  mergers amongst the dynamically formed population  (see also Table~\ref{tab:finalstate}). Moreover, it has little effect on the  time when the first and subsequent mergers among the dynamically formed population happen. 

With Fig.~\ref{fig:RadvsTime} we show both variations of models Z3-M10-D3 (upper panels) and Z1-M10-D3 (lower panels). Here we plot the radial position of every BBH in the simulations, distinguishing between dynamically formed BBHs and those from the primordial population. For the ejected BBH population we extend their tracks using equation~(\ref{eq:GWTime}) either up to a Hubble time ($t_{\rm H}$) or until they reach coalescence. This allows us to include BBH mergers among the ejected population that occur after the end  of the $N$-body simulation but within $t_{\rm H}$.   

 Considering the evolution of primordial BBHs, we see that in both models there are many BBHs which are ejected from the cluster  shortly after being formed. Of these BBHs almost all of them go on to merge within $t_{\rm H}$. As discussed previously, dynamical encounters are a key mechanism for ejection of BHs and BBHs from a cluster; however, we find that these BBHs are  ejected by the natal kick on one of the binary components. This early ejection from the cluster by a stellar evolution mechanism, provides a sub-population of merging BBHs
 that likely show negligible effect from dynamics. Later in the lifetime of the cluster, we also find ejections of primordial BBHs, some of which also merge.  These later ejections are  due to  dynamical encounters in the cluster.

As can be seen in Table~\ref{tab:finalstate}, the primordial binary mergers are split roughly evenly between in-cluster mergers and ejected mergers in all models. In contrast, most  dynamically formed BBHs  merge inside the cluster. Moreover, we find that primordial BBHs are always the dominant source of mergers. We further discuss in-cluster and ejected mergers later in Section~\ref{sec:inclvseject}.

\begin{figure*}
    \centering  
\includegraphics[width=2\columnwidth]{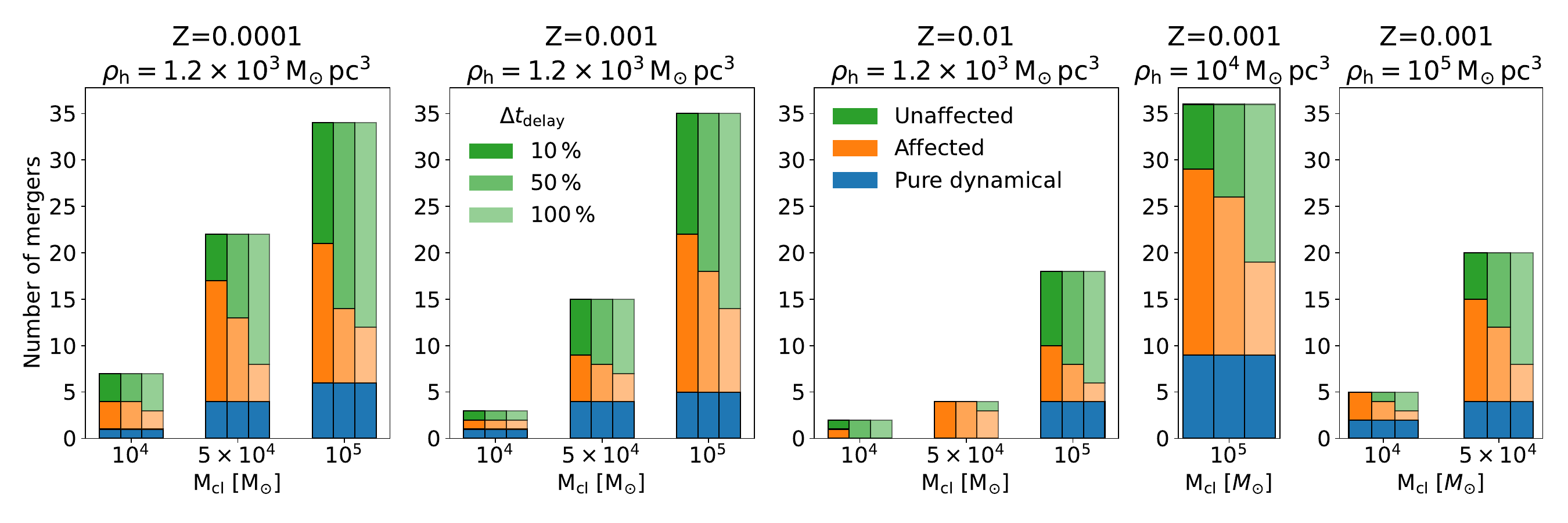}
    \caption{The number of mergers in our cluster models, split between the dynamical population and the primordial population. The primordial population is further split into "\textit{affected}" and "\textit{unaffected}" binaries according to the fractional change in delay time from a purely isolated evolution. We show these counts for three cut-off fractional change values in the delay time: 10\%, 50\% and 100\%.}
    \label{fig:affctBins}
\end{figure*}

\subsection{Effect of dynamics on the number of BBH mergers}\label{sec:Dynms}

We have shown that roughly half of the mergers amongst the primordial BBH population come from BBHs ejected by natal kicks shortly after formation, and thus likely contain no imprint from the dynamics of the cluster. On the other hand, the 
remaining BBH mergers that are produced by binaries that remain bound to the cluster after the SN kicks can show some imprint of the dynamical environment in which they evolved. To investigate the extent to which the cluster has affected the population of primordial BBH mergers, we take the primordial binaries in each cluster and evolve them using the stand-alone $\tt BSE$ code. The binary populations are simulated until merger and then the merger times are compared against the same specific binary in the cluster simulations.

Since the stellar evolution prescriptions are chosen identically to the corresponding full cluster simulation, any difference in the merger time for a specific binary can be attributed to the dynamics of the cluster altering the binary parameters. However, we do not account for the randomness of the drawn natal kick magnitude and direction in our comparison and thus our results represent an upper estimate for the number of BBHs that are affected by dynamics.

When comparing the merger time for a specific binary, we define the fractional difference in merger time:
\begin{equation}
\Delta t_{\mathrm{delay}} = \frac{\left| t_{\mathrm{d, \,isol}} - t_{\mathrm{d, \,cluster}} \right|}{t_{\mathrm{d, \,isol}}},
    \label{eq:fractdelay}
\end{equation}
where $t_{\mathrm{d, \,isol}}$ is the merger time from the isolated binary simulation and $t_{\mathrm{d, \,cluster}}$ is the merger time in the cluster simulation. Note that here we are taking the absolute value of the  difference since the dynamics of the cluster can either aid the merger of the binary or hinder it. We then make a choice for the boundary value of the fractional difference; if the computed fractional difference for a given BBH is larger than the cut-off value, we categorise the BBH as \textit{affected} by the cluster. 
 We choose three cut-off values, $10\%,\,50\%$ and $100\%$ -- a  $100\%$ difference  occurs when the inclusion in a cluster environment has changed a BBHs delay time by a factor of 2.
Fig.~\ref{fig:affctBins} shows the results of this comparison for a selection of our models across varying initial cluster mass, density and metallicity. We find that  $\lesssim 20\%$  (for $100\%$ change in delay time) of the primordial BBH mergers are characterised as affected by dynamics. Based on our theoretical understanding of BBH dynamics in clusters \citep{breen_dynamical_2013, rodriguez_post-newtonian_2018}, we would expect that as the mass and density of the cluster are increased the effect of dynamical encounters will become  more important.  On the contrary, our results  appear to be  mostly independent of cluster mass, density and metallicity, in the sense that the fraction of BBH mergers that have been significantly affected by dynamics remains similar across all models. For $M_{\rm cl}=10^5\,{\rm M_\odot}$ and $Z=0.001$, the fraction of mergers among the primordial binary population that have been significantly affected by dynamics ($100\%$ variation in delay time) is $0.2$ for $\rho_{\rm h}=1.2\times 10^{3}\,\density$ and $0.15$ for $\rho_{\rm h}=10^{5}\,\density$. For $\rho_{\rm h}=1.2\times 10^{3}\,\density$ and $Z=0.0001$, the fraction of affected mergers is $0.3$ for $M_{\rm cl}=10^4\,{\rm M_\odot}$ and $0.14$ for $M_{\rm cl}=10^5\,{\rm M_\odot}$. However, if we consider a 10\% variation in delay time; we find that the majority of BBHs in high density clusters ($\rho_{\rm h} = 10^{4}~\rm M_{\sun}~pc^{-3}$ and $\rho_{\rm h} = 10^{5}~\rm M_{\sun}~pc^{-3}$) are characterised as affected, compared with far fewer (typically around half) in  lower density clusters. This suggests more dynamic activity in regards to weaker interactions in the high density clusters which produce only a small change in the properties of the primordial BBHS.

We have seen that our clusters contain a population of primordial BBHs which are ejected early on due to the large component natal kicks. Since these kicks are drawn randomly in both the stand-alone $\tt BSE$ code and $\tt PeTar$, they could be the cause of the \textit{affected} binaries we find in Fig.~\ref{fig:affctBins}. We investigate this possibility by performing the same analysis, excluding this group of escaped BBHs. If the kicks were the most important factor, we should expect the fraction of affected binaries to decrease significantly in this new analysis, since the ejected population are the binaries that receive the largest kicks. We find that there is effectively no difference in the fraction of affected mergers from the primordial binary population. Comparing to the values stated above: for $M_{\rm cl}=10^5~\rm M_{\sun}$ and $Z=0.001$ we find the fraction of affected mergers is $0.38$ at $\rho_{\rm h}=1.2\times10^{3}~\rm\density$ and $0$ at $\rho_{\rm h}=10^5~\rm\density$. For $\rho_{\rm h}=1.2\times 10^{3}\,\density$ and $Z=0.0001$, the fraction of affected mergers is $0.4$ for $M_{\rm cl}=10^4\,{\rm M_\odot}$ and $0.15$ for $M_{\rm cl}=10^5\,{\rm M_\odot}$. The small changes in the fraction of affected binaries compared to our former analysis (Fig.~\ref{fig:affctBins}) likely implies that the larger effect on the binary properties is due to dynamical encounters. 

\begin{figure*}
    \centering
    \includegraphics[width=\columnwidth]{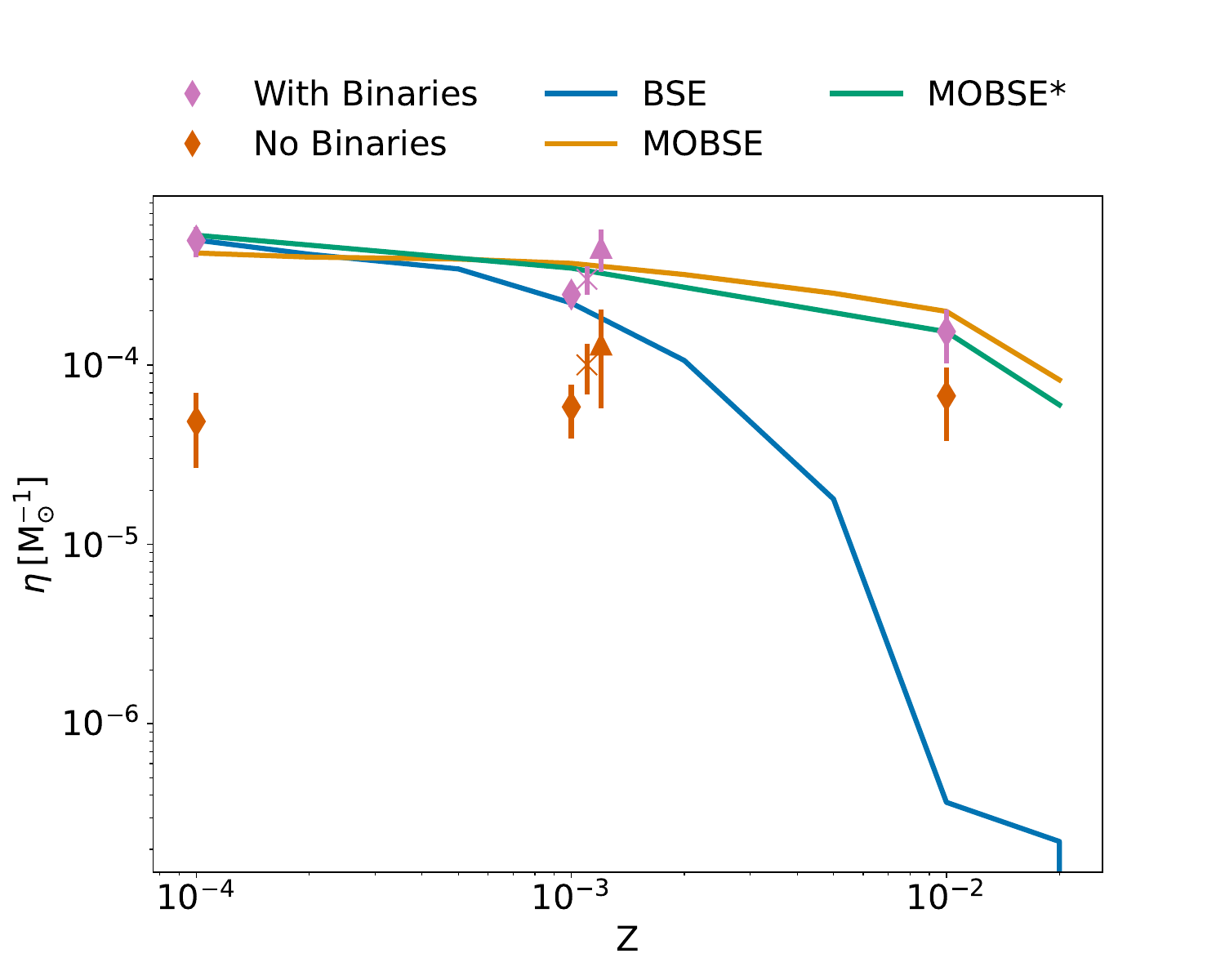}\includegraphics[width=\columnwidth]{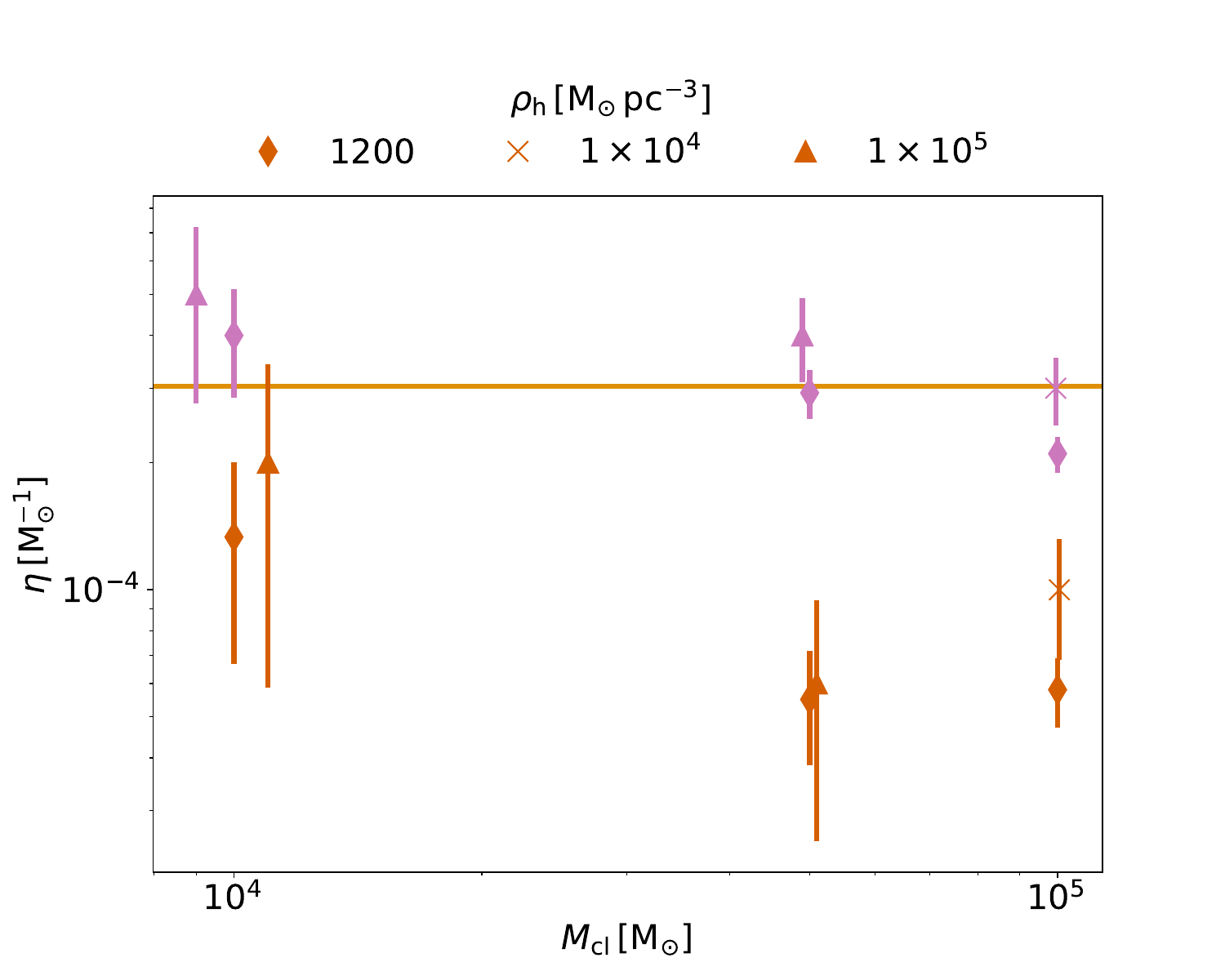}
    \caption{The merger efficiency as a function of metallicity (left panel) and initial cluster mass (right panel). We distinguish between the clusters with a primordial population and without by purple and green markers respectively, whilst the initial cluster density is shown by marker type. We also over plot the merger efficiency from the stellar evolution codes $\tt BSE$ and $\tt MOBSE$ shown by orange and blue lines. We found a large discrepancy between these two codes at high metallicity owing to the treatment of Hertzsprung gap stars during common envelope evolution. We show that resolving this discrepancy gives an adjusted $\tt MOBSE$* relationship in red which is more consistent with the $\tt BSE$ results.}
    \label{fig:mergeEff}
\end{figure*}

To determine the dependence of the number of BBH mergers on the cluster properties and how this number is affected by dynamical encounters, we compute the cluster merger efficiency ($\eta$).
\begin{equation}
    \eta = \frac{N_{\mathrm{merge}}}{M_{\mathrm{cl}}},
    \label{eq:mergeEff}
\end{equation}
where $N_{\mathrm{merge}}$ is the number of mergers in the cluster. To compare against the expected efficiency from the isolated channel, we also simulate a population of binaries using two different binary evolution codes, $\tt BSE$ and $\tt MOBSE$ \citep{giacobbo_merging_2018, giacobbo_progenitors_2018}. The initial binary populations in these latter models were the same as the initial binaries in the cluster simulations. 
The left (right) panel of Fig.~\ref{fig:mergeEff} shows  $\eta$ as a function of metallicity (cluster mass), where we differentiate between clusters with primordial binaries and those without. 
We see that the merger efficiency of clusters with a primordial binary population broadly follows the same relationship as the same population simulated in isolation using $\tt BSE$. We conclude that in our models, dynamical encounters have a small effect on the BBH merger rate.

In the left  panel of Fig.~\ref{fig:mergeEff} we see that the results obtained with $\tt MOBSE$ show a large disagreement with both the isolated $\tt BSE$ results and the cluster simulation results. This  discrepancy is due to the treatment of Hertzsprung gap (HG) stars during a common envelope (CE) evolution phase. 
$\tt BSE$  allows for the possibility of binary survival when the CE is initiated by a star crossing the
HG.
In the standard version of $\tt MOBSE$, instead, when a HG star enters a CE phase as a donor star, it is assumed that the stars  merge. This assumption leads to a small number of systems surviving a CE phase at high metallicity -- the rapid expansion of metal-rich stars in the HG that initiate a CE leads to stellar mergers due to the absence of a well-developed core-envelope structure.
In contrast, metal-poor stars remain relatively compact in the HG but expand more significantly in the subsequent stellar evolution phases. These facts fully explain the large difference in the merger efficiency obtained with $\tt BSE$ and $\tt MOBSE$ at $z>0.001$. To further illustrate this, we evolve the same binary population 
with $\tt MOBSE$, but now allowing the binaries to survive a CE phase that occurs during the HG (indicated as $\tt MOBSE^*$ in Fig.~\ref{fig:mergeEff}). As expected, these new simulations  recover a similar  merger efficiency as found with $\tt BSE$ and the cluster models. Finally, in Fig.~\ref{fig:mergeEff} we compare the merger efficiency  between cluster models with and without primordial binaries and find that the former always show a high merger efficiency. The merging efficiency for each cluster simulation is summarised in Table~\ref{tab:finalstate}. 

The results discussed in this section and illustrated in  Fig.~\ref{fig:affctBins} and Fig.~\ref{fig:mergeEff} 
lead to the following conclusions: (i)  most  BBH mergers found in our models  with a primordial binary population are not assembled dynamically, although a significant fraction of them are still  affected by dynamical encounters (see Fig.~\ref{fig:affctBins}); (ii) the merger efficiency of our cluster models  with an initial binary population  is not significantly increased by dynamical encounters. This can be seen in Fig.~\ref{fig:mergeEff} by comparing the value of $\eta$ for the isolated binary models and the cluster models; and (iii) the role of dynamical encounters in enhancing the merger rate of BBHs  depends on the stellar evolution prescriptions used. Specifically, if  stars are allowed to merge during a CE phase occurring when the donor is on the HG, dynamically assembled BBHs will  dominate for  $Z>0.001$, while they remain a subdominant population at lower metallicities.

\subsubsection{Higher multiplicity systems}
As previously shown, a fraction of the primordial BBH mergers are affected by the dynamics, and almost all dynamical BBH mergers occur in-cluster. Thus, it is reasonable to assume that at least some of these merging systems may participate to even higher multiplicity interactions involving triples, quadruples etc. We investigate this and find that $\lesssim 10\%$ of all primordial BBHs mergers occur in a triple BH system, with the rest occurring in binaries. On the other hand, the dynamically formed BBH mergers showed roughly equal number of mergers as a part of a stable triple and as a binary. We also find a small fraction of dynamical BBHs merging whilst part of a quadruple BH system. No higher multiplicity systems were found to contain merging BBHs. 

\begin{figure}
    \centering
    \includegraphics[width=\columnwidth]{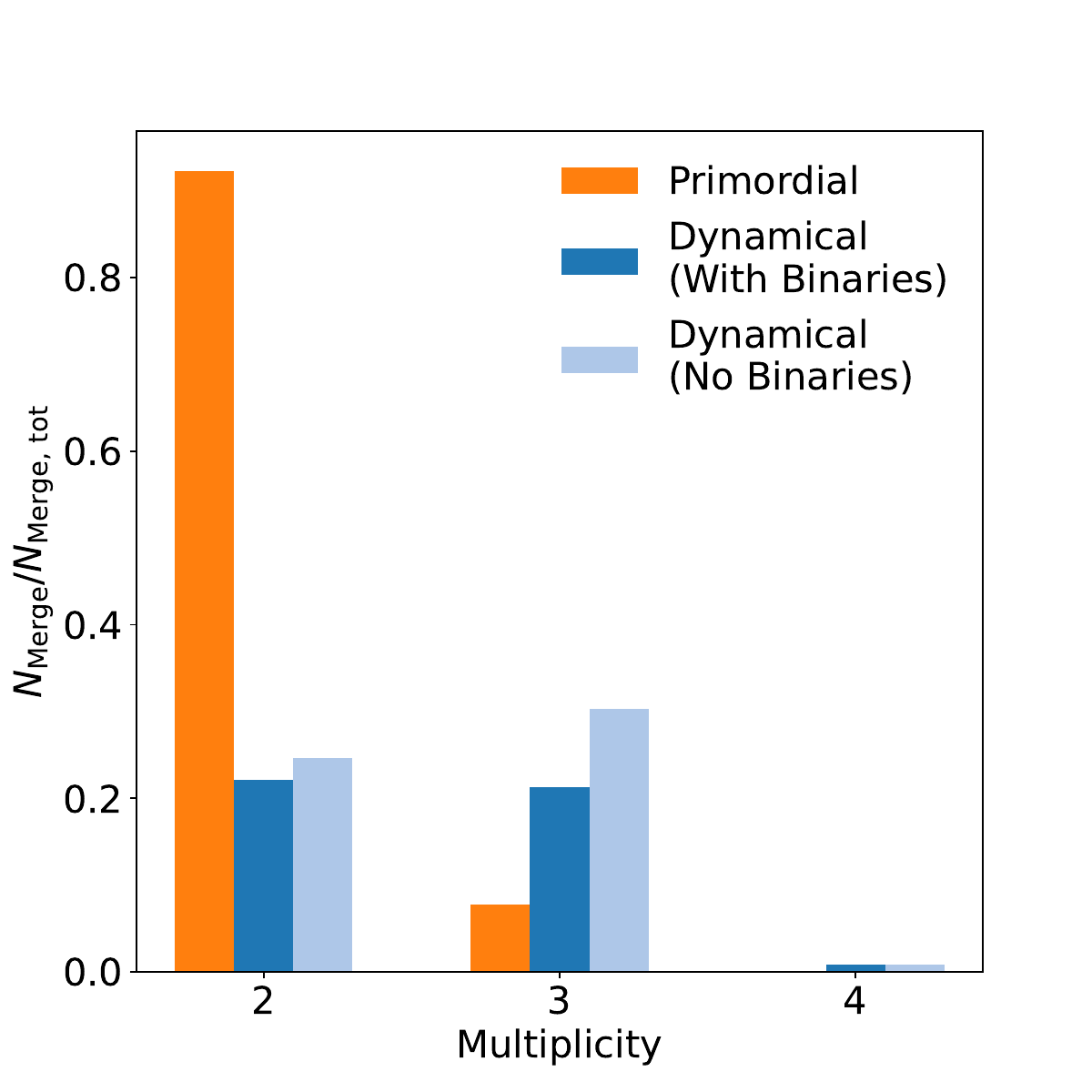}
    \caption{Fraction of mergers that are found in stable triples and quadruples just before merger. We distinguish between mergers from the primordial binary population and the dynamically formed population, as well as further splitting the dynamically formed binaries based on whether the cluster contained a primordial binary population. Higher multiplicity systems were searched also, however no mergers were found within them.}
    \label{fig:multiplicity}
\end{figure}

Fig.~\ref{fig:multiplicity} shows the fraction of merging BBHs that are found in each type of system, and we have further split up the dynamical BBH mergers in those from clusters with a primordial binary population and those without. We see that in clusters without primordial binaries the dynamical BBH mergers occur slightly more frequently within stable triple systems than  in binaries.

\begin{figure}
    \centering
    \includegraphics[width=\columnwidth]{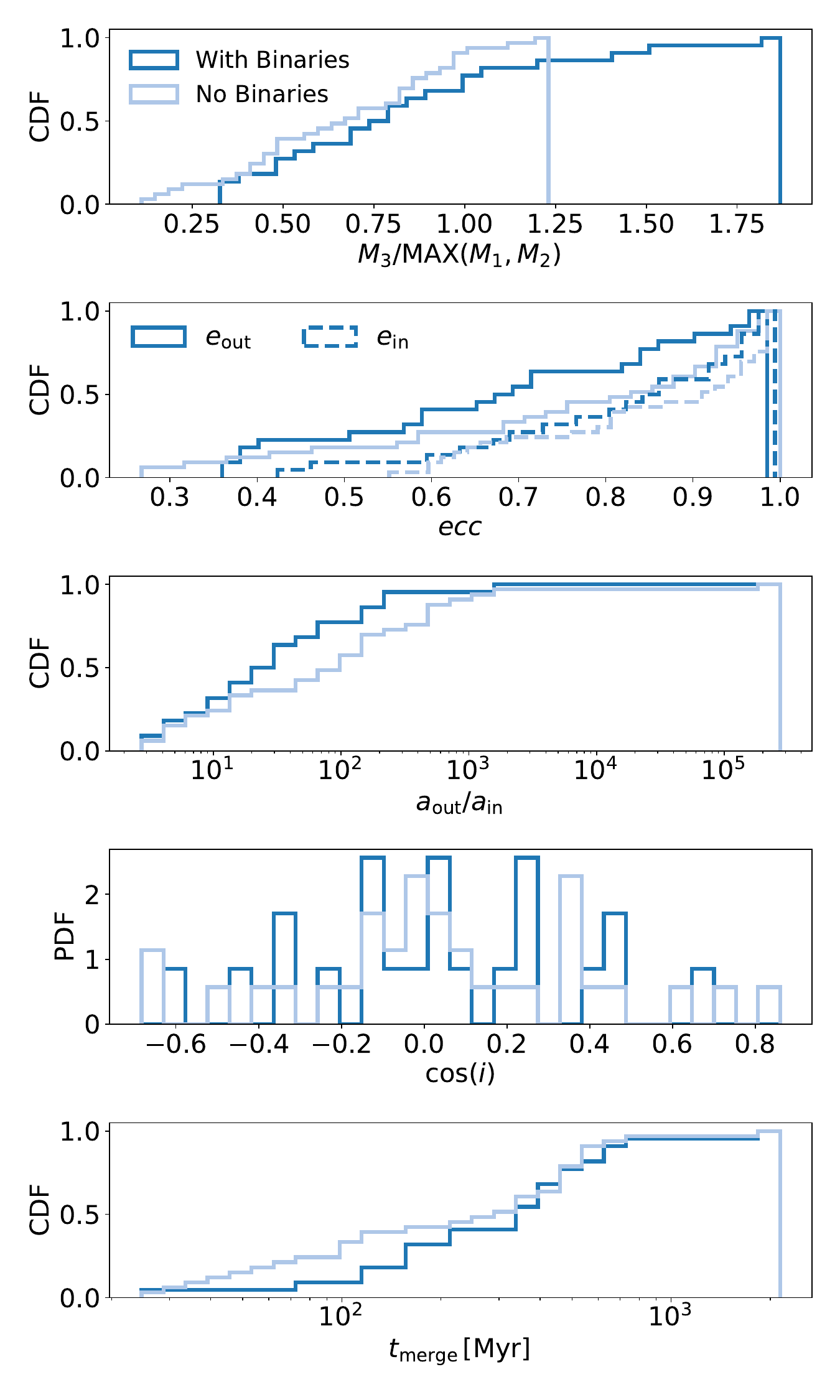}
    \caption{The orbital distributions for the stable triple systems that contain a dynamically formed BBH which merges within $t_{\mathrm{H}}$. We split the dynamically formed BBHs into the population coming from clusters with a primordial binary population and those without.}
    \label{fig:TripDists}
\end{figure}

The orbital properties of the stable triple systems that are formed dynamically in our models are shown in  Fig.~\ref{fig:TripDists}.  
These are obtained from the last $N$-body snapshot in which  the binary was still present in the simulation. 
At this time, the exact merger time due to GW energy loss varied slightly between binaries, but it was always $\lesssim 1\,\mathrm{Myr}$. 
We consider the distribution of $M_3/\max(M_1,M_1)$, where $M_3$ is the mass of the tertiary object on a outer orbit  with semi-major axis $a_{\rm out}$ and eccentricity $e_{\rm out}$. The relative inclination between the inner and outer binary orbits is indicated as $i$. The analysis shown in Fig.~\ref{fig:TripDists} helps to understand whether the presence of a tertiary BH can affect the inner BBH evolution -- lighter and more distant tertiary companions will have a smaller effect on the evolution of the binary.
Interestingly, from the top panel of Fig.~\ref{fig:TripDists} we see that the most massive object of the triple is in most cases one of the binary components. We also find that both inner and outer orbits often have a significant eccentricity.
The relatively small values of the ratio $a_{\rm out}/a_{\rm in}\lesssim 10^3$ indicate that at least in some cases we might expect the tertiary to have an  effect on the evolution of the binary. However, we note that in order to address the impact on the BBH evolution, one should consider a  more detailed analysis,  taking into account  relativistic effects acting on the the inner binary orbit \citep[e.g.,][]{ford_secular_2000, blaes_kozai_2002}.

\subsection{Effect of dynamics on the properties of BBH mergers}\label{sec:mergers}

\begin{figure*}
    \centering
    \includegraphics[width=1.8\columnwidth]{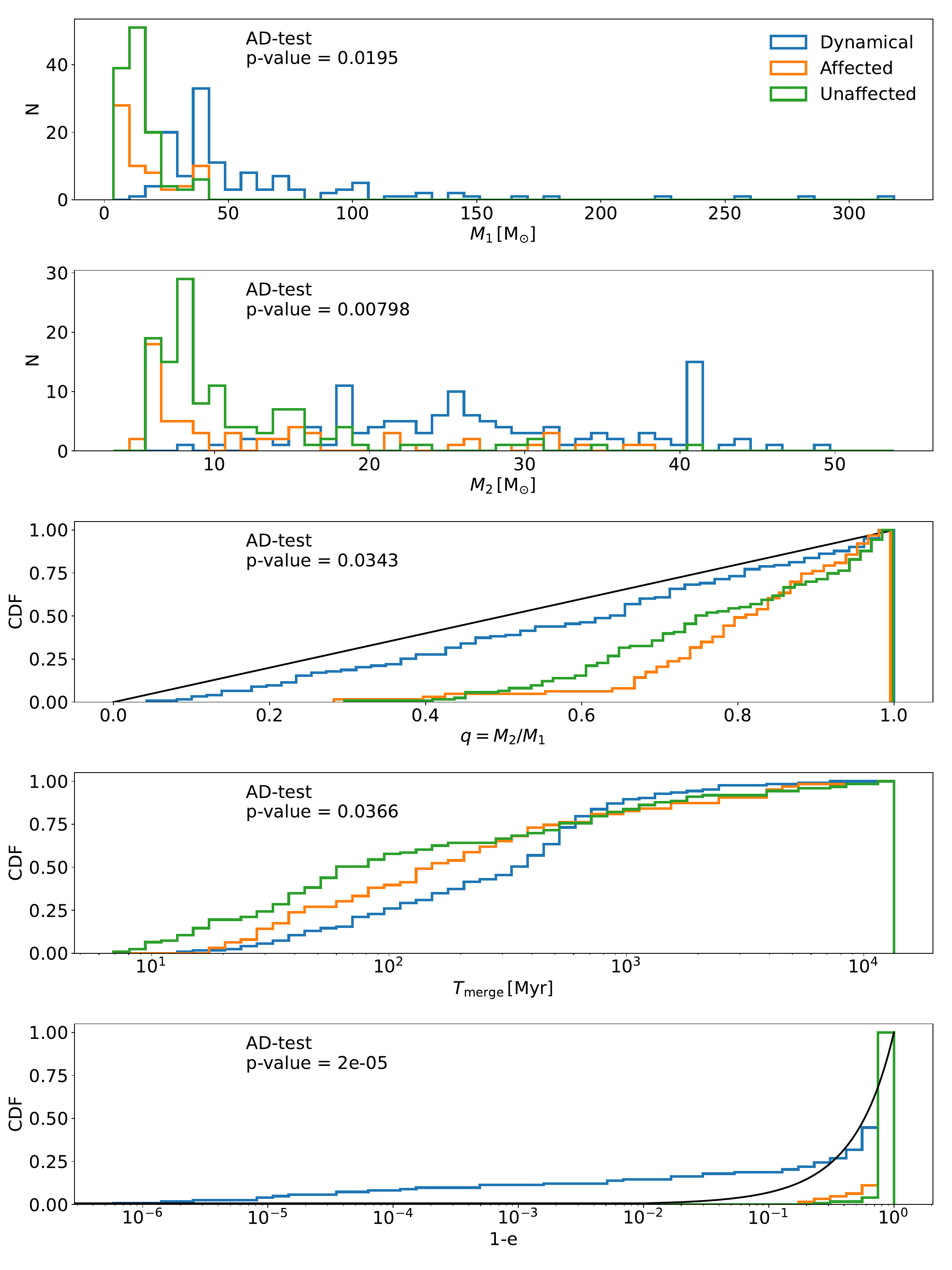}
    \caption{Distributions for the three merging populations - dynamically formed BBHs, affected primordial BBHs and unaffected primordial BBHs, from all simulations. Here we have taken the cut-off fractional change defining affected binaries as 50\%. We note that for each distribution we perform a K-sample Anderson-Darling \citep{scholz_k-sample_1987} test between the unaffected and affected populations, the P-value of the tests are shown on the corresponding panel. For the mass ratio and eccentricity panels, we plot the reference distributions $U(0,1)$ and $f(<e)\propto e^2$ respectively.}
    \label{fig:BBHDists}
\end{figure*}

We have shown in the previous sections that a significant fraction of primordial BBHs can be  affected by dynamical encounters in the cluster, and that despite this the number of mergers is consistent with the same binaries evolving in isolation. However, it is likely that the binary orbital parameters of this affected population differ compared to the unaffected BBHs. To investigate this, we compute the distributions of component masses, mass ratio, merger time and eccentricity. We split up the mergers in dynamically formed BBHs, primordial BBHs that are affected by the cluster (where we take affected BBHs as those with $\Delta t_{\mathrm{delay}}\leq50\%$) and primordial BBHs unaffected by the cluster. These distributions are plotted in Fig.~\ref{fig:BBHDists}. 

For each parameter we take the affected and unaffected population and perform an Anderson-Darling (AD) k-sample test \citep{scholz_k-sample_1987} to test whether the samples from the two populations are drawn from the same distribution. We opt for the AD test since this gives more weight to the tails of the distribution compared to other tests (such as the Kolmogorov-Smirnov test). The p-value from this test is shown on each of the distribution plots. Our results show that at the 95\% level, the unaffected and affected population are sampled from different distributions for every parameter. Further, the secondary mass and eccentricity samples for the two populations are sufficiently different at the 99\% level. 

The bottom panel in Fig.~\ref{fig:BBHDists} shows the eccentricity measured at the moment the binary decouples from the cluster dynamics, which is determined as described in Section~\ref{sec:methods}. The distribution shows that the affected BBHs are typically more eccentric than the unaffected BBHs. This is likely due to dynamical encounters within the cluster. It is typically assumed that on average eccentricity induced through single-binary encounters should follow a thermal distribution (shown by the black line in the plot) \citep{heggie_binary_1975}. However, we see that the affected BBH population does not follow this relation. 
A reasonable explanation for this is that dynamical encounters did not have enough time to fully thermalise the distribution before GW energy loss  leads to orbital circularisation. However, the effect of encounters is still significant enough that the eccentricity distribution of the affected population can be distinguished from the unaffected one. As a caveat to this analysis, we stress that these differences might at least partly explained by the different natal kick magnitude and direction that are drawn randomly from the assumed distributions. However, following the analysis in Section~\ref{sec:Dynms} this is unlikely.

We now focus on the dynamical BBH population and their orbital parameters. 
Several important results emerge from this analysis. First the mass-ratio distribution of merging BBHs appears to contain more asymmetric binaries than the primordial BBH mergers, and it is nearly uniform between 0 and 1.
Both of these results are somewhat surprising since three-body encounters are expected to favour the formation of binaries with nearly equal mass components~\citep{ rodriguez_binary_2016, park2017black}. Moreover, we find that the BBH eccentricity distribution is clearly over-thermal, in the sense that it contains more eccentric binaries than  $N(<e)\propto e^2$; about $20\%$ of the BBHs have $1-e\lesssim 10^{-2}$. This is not surprising, however, because we are  only considering  those BBHs that merge within $t_{\rm H}$, which naturally favours the BBHs with the highest eccentricities due to the strong dependence of the GW merger timescale on  $e$.

\begin{figure}
    \centering
    \includegraphics[width=1\columnwidth]{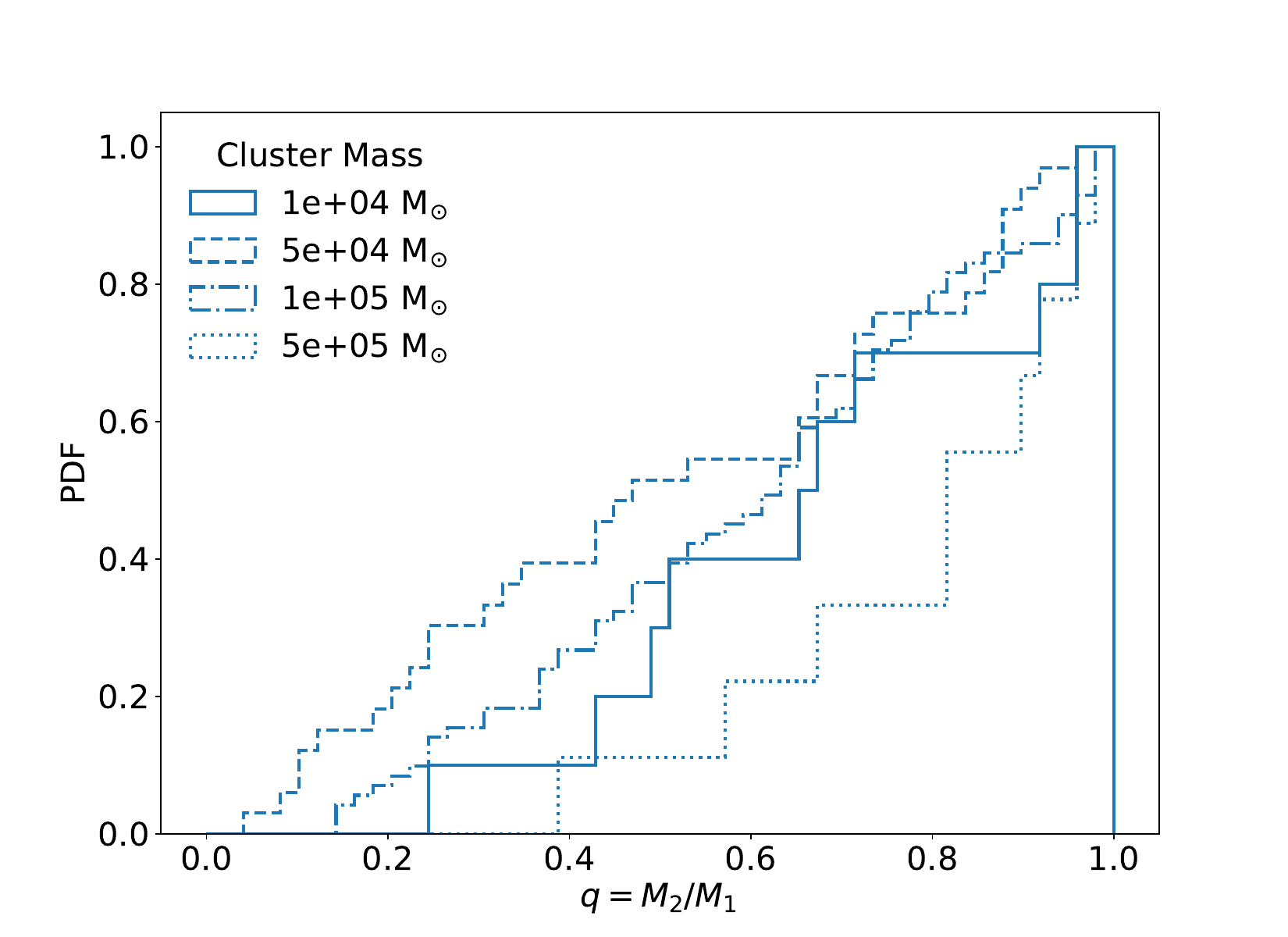}
    \caption{The cumulative distribution of the mass ratio $(q)$ for dynamically formed merging BBHs. We show the combined distribution from all of our simulations, as well as the distribution for each initial cluster mass.}
    \label{fig:qCDF}
\end{figure}

\begin{figure*}
    \centering
    \includegraphics[width=1.8\columnwidth]{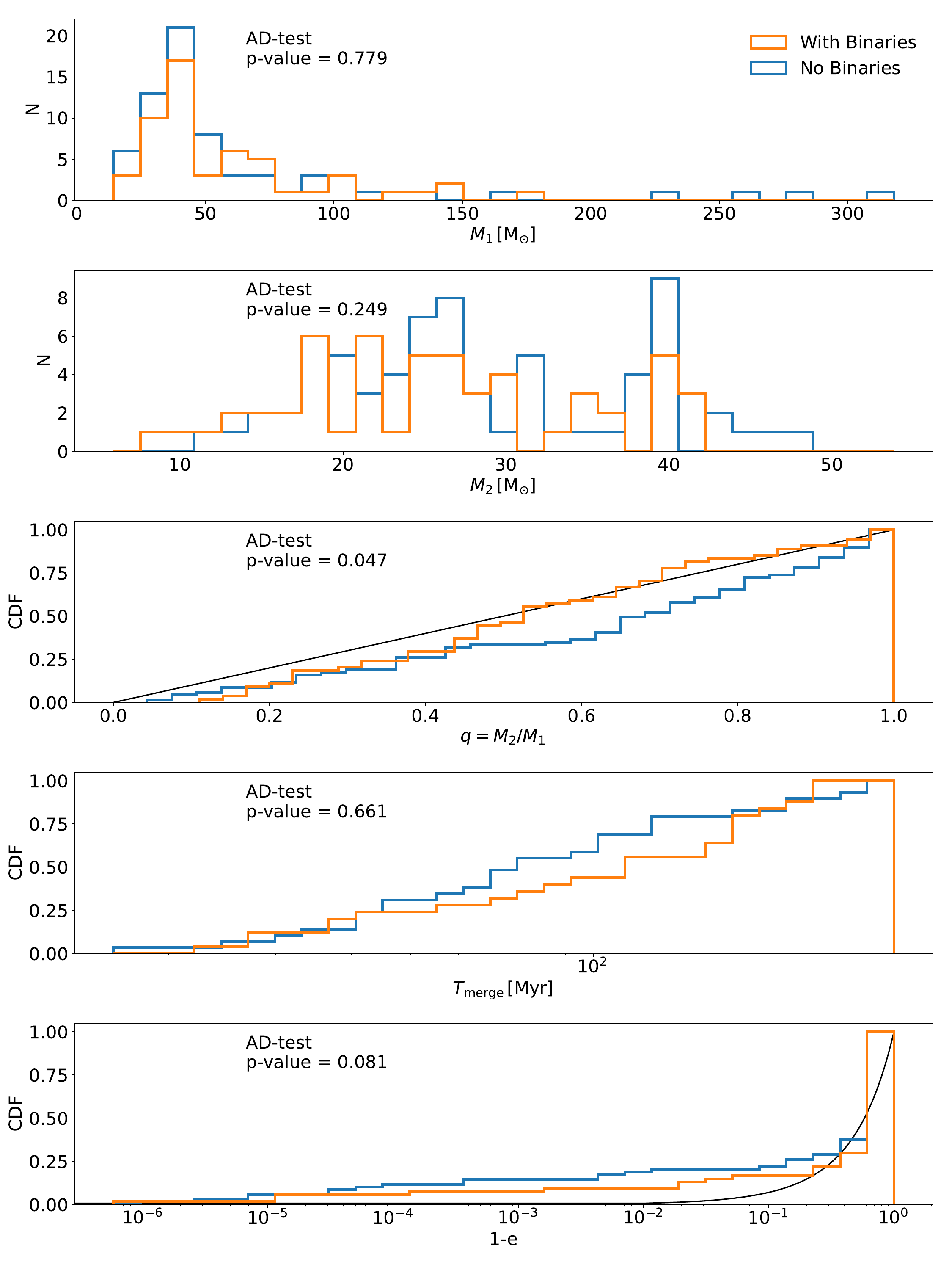}
    \caption{Distributions of dynamical formed BBHs from all clusters with and without a primordial binary population. We note that for each distribution we perform a two sample Anderson-Darling \citep{scholz_k-sample_1987} test between the two populations, the P-value of the tests are shown on the corresponding panel. mass ratio and 1-eccentricity we plot the reference distributions $U(0,1)$ and $f(<e)\propto e^2$
    respectively.}
    \label{fig:compareDynmMergers}
\end{figure*}

In addition, we investigate whether the presence of the primordial population affects the properties of the dynamically formed BBHs that go on to merge. We therefore split the dynamical BBH population depending on whether its host cluster contained a primordial binary population and again perform an AD k-sample test \citep{scholz_k-sample_1987} between these two populations. Fig.~\ref{fig:compareDynmMergers} shows the results of these comparisons for the same parameters as in Fig.~\ref{fig:BBHDists}. Our tests suggest for each of the parameters the two populations of dynamical BBHs are drawn from the same distribution, and so the presence of the primordial population does not affect the merging dynamical BBHs.

One of our cluster variations (Z2-M10-D3-L*) is initialised with a \citet{duquennoy_multiplicity_1991} binary period distribution. Although not the focus of this work, we investigated whether this choice led to any differences in the BBH properties. We found little to no difference in the BBH properties compared to clusters initialised with a \citet{sana_binary_2012} period distribution.

\section{in-cluster vs Ejected Mergers}\label{sec:inclvseject}

$N$-body simulations provide a detailed understanding of the evolution of stellar populations within a cluster, requiring minimal assumptions about the properties and dynamical evolution of the BHs. As such, they represent the most reliable approach for advancing our understanding of BBH formation in dense stellar environments. However, due to the substantial computational demands of $N$-body simulations, approximate methods--such as those relying on Monte Carlo techniques and semi-analytical codes-- are often used. These methods typically involve the following two assumptions (among other ones): (i) the only important interactions in the cluster core are those close interactions that involve binaries and single objects, i.e., interactions between  binaries and singles and interactions between two binaries; (ii) these  interactions are assumed to be strong interactions, which lead to a either a direct or a resonant encounter between the three BHs.
Assumption (i) means that interactions involving higher multiplicity systems such as stable triples and quadruples are often neglected. Assumption (ii) implies that the effect of soft interactions --where the closest approach of the third BH to the binary is larger than $\sim 2$ times the binary semi-major axis-- is also neglected. 

Given these assumptions, the number of merging BBHs can be approximately derived. Moreover, one can derive the number of mergers that occur inside the cluster $vs$ those that occur after being dynamically ejected from it. This distinction is a key to  a full characterisation of the merging BBH population. For example, a fraction of in-cluster mergers  are expected to have a residual eccentricity within the frequency band of current GW detectors. Moreover,  mergers among the ejected and the in-cluster populations will have different redshift distributions,
as in-cluster mergers occur earlier on during the evolution of the cluster. 

In this section, we begin by reviewing the theory framework for the formation and evolution of BBHs in star clusters. We then compare these theoretical predictions with the outcomes of our cluster models, providing a  testbed to evaluate and refine current theories.

\subsection{Theory}\label{theory}
Here we follow \cite{samsing_eccentric_2018} and \cite{antonini_population_2020} to describe the evolution of BBHs due to binary-single interactions.
We assume that after a hard BBH is formed 
in the core of a star cluster, it experiences a sequence
of binary-single interactions with single BHs and stars. 
During  a single interaction with a cluster member of mass $M_3$
the semi-major axis of the binary decreases from $a$
to $\epsilon a$. Then energy and momentum conservation imply
that the binary   experiences a  recoil kick 
 $v_{\rm bin}^2=G\mu {M_3\over
  {M_{123}}} \left(1/\epsilon a-1/a\right) \approx 0.2 G\mu {M_3\over
  {M_{123}}} {q_3/a}$, where $\mu=M_1M_2/M$, $M_{123}=M_1+M_2+M_3$,
$q_3=M_3/(M_1+M_2)$, and we have assumed that in the interaction
the binding energy of the binary increases by a fixed fraction
$\approx 0.2$,
 i.e., $\epsilon \approx 1/(1+0.2)$ \citep[e.g.,][]{quinlan_dynamical_1996, coleman_miller_production_2002, antonini_merging_2016}.
 Setting $v_{\rm bin}= v_{\rm esc}$, with $v_{\rm esc}$ the escape velocity from the cluster,
 we  obtain the limiting semi-major axis below which a three
body interaction will eject the binary from the cluster:
\begin{eqnarray}\label{aej}
a_{\rm ej}&= &\left({1\over \epsilon}-1\right)G {M_1M_2\over M_{123}} {{q_3}}/{v_{\rm esc}^{2}} .
\end{eqnarray}
We have two possibilities, either the binary reaches 
$a_{\rm ej}$ and it is ejected from the cluster or it mergers  before reaching $a_{\rm ej}$.

The total probability that a binary merges in between
two consecutive binary-single interactions  is
obtained by integrating the differential merger probability 
per binary-single encounter, $d\mathcal{P}_{\rm GW}= P_{\rm GW}dN_{3}$, over the
total number  of binary-single interactions experienced by the binary \citep{samsing_eccentric_2018}.
Noting that  $da/dN_{\rm 3}=(\epsilon-1)a$, this leads to 
\begin{eqnarray}\label{Prob}
\mathcal{P}_{\rm GW}(a_{\rm ej}) & =&  
\int_{a_{\rm h}}^{a_{\rm ej}} 
{1\over \epsilon-1}
{\ell_{\rm GW}^2 } {da\over a} \approx
{7\over 10}{1\over 1-\epsilon}\ell_{\rm GW}^2(a_{\rm ej}),
 \end{eqnarray}
 where 
$\ell_{\rm GW}$ is the value of $\ell=(1-e^2)^{1/2}$ below which the evolution of the binary becomes dominated by GW energy loss -- we are assuming here that the binary  receives a large angular momentum kick such that the phase space is stochastically scanned and  uniformly covered by
the periapsis values.
\cite{antonini_population_2020}
showed that 
\begin{eqnarray}\label{gw}
\ell_{\rm GW}\simeq
1.3\left[
{G^4\left(M_1M_2\right)^2(M_1+M_2)
\over
c^5} {t_{\rm rh} \over \zeta |E|}
\right]^{1/7}a^{-5/7},
\end{eqnarray} 
where $\zeta \simeq 0.1$ and $E=-0.2GM_{\rm cl}^2/r_{\rm h}$, with $r_{\rm h}$ the half-mass radius of the cluster.

 The total probability  that a BBH will merge outside its parent cluster
is given by the product of the probability that the binary reaches $a_{\rm ej}$ and the probability that it merges after being ejected \citep{antonini_population_2020}
\begin{equation}\label{pex}
\mathcal{P}_{\rm ex}(a_{\rm ej})=\left(1-\mathcal{P}_{\rm GW}(a_{\rm ej})\right)P_{\rm ex}\ ,
\end{equation}
where $P_{\rm ex}$ is 
the  probability 
that an ejected binary  merges on a timescale shorter  than $t_{\rm H}$: 
\begin{equation}
P_{\rm ex}(a_{\rm ej})=\ell_{\rm H}^2(a_{\rm ej}),
\end{equation}
and 
\begin{eqnarray}
\ell_{\rm H}\simeq 1.8\left[\frac{G^3M_1M_2(M_1+M_2)}{c^5} t_{\rm H}
\right]^{1/7}a^{-4/7}.
\end{eqnarray}
Thus, for $\ell < \ell_{\rm H}$, an ejected binary merges in less than $t_{\rm H}$.

\begin{figure}
    \centering    
    \includegraphics[width=\columnwidth]{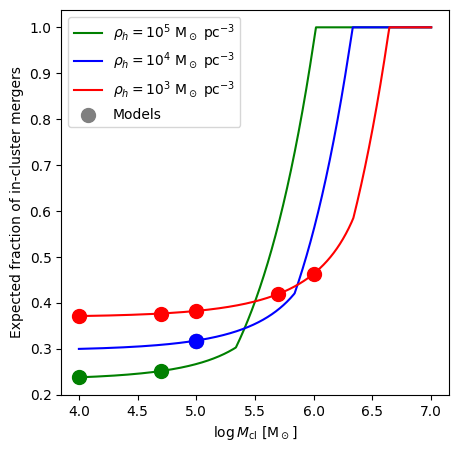}
    \caption{Expected fraction of in-cluster mergers
    as a function of cluster mass and density obtained 
   from equation~(\ref{fin_th}). The filled circles show the predicted fraction of in-cluster mergers for the initial values
    of $M_{\rm cl}$ and $\rho_{\rm h}$ we used in the $N$-body models.}
    \label{fig:predictedFin}
\end{figure}

From the above considerations it follows that the fraction of 
in-cluster mergers to the total population of BBH mergers produced by a cluster
is 
\begin{equation}\label{fin_th}
\mathcal{F}_{\rm GW}=
{\mathcal{P}_{\rm GW}\over \mathcal{P}_{\rm GW} +\mathcal{P}_{\rm ex}} \ .
\end{equation}
In Fig.~\ref{fig:predictedFin} we plot this quantity as a function of cluster mass and for different values of $\rho_{\rm h}$. For the cluster models we considered, we should expect that the fraction of
merges that are produced inside the cluster varies between $\approx 0.25$ and $0.45$. These fractions appear to be consistent with those obtained  in previous work.
For example, \citet{rodriguez_post-newtonian_2018-1} find that $55\%$ of the mergers in their Monte Carlo models occur when the binary is still bound to its parent cluster.

\subsection{Comparison to $N$-body models }

We investigate the fraction of in-cluster mergers in our simulations for both primordial and dynamically formed BBH populations. To make a comparisons against theoretical studies, here we simulate the same clusters using the fast cluster population model code $\tt cBHBd$ \citep{antonini_population_2020}.
This code assumes no primordial binaries, and only considers BH-BH interactions through binary-single encounters. 
The basic theoretical framework is the one described above in Section~\ref{theory}.
Using $\tt cBHBd$ we produce 1000 realisations of each of our cluster initial conditions (see Table~\ref{tab:initCond}), we then find the average number of in-cluster and ejected mergers as well as the average in-cluster fraction.

\begin{figure*}
    \centering
    \includegraphics[width=\columnwidth]{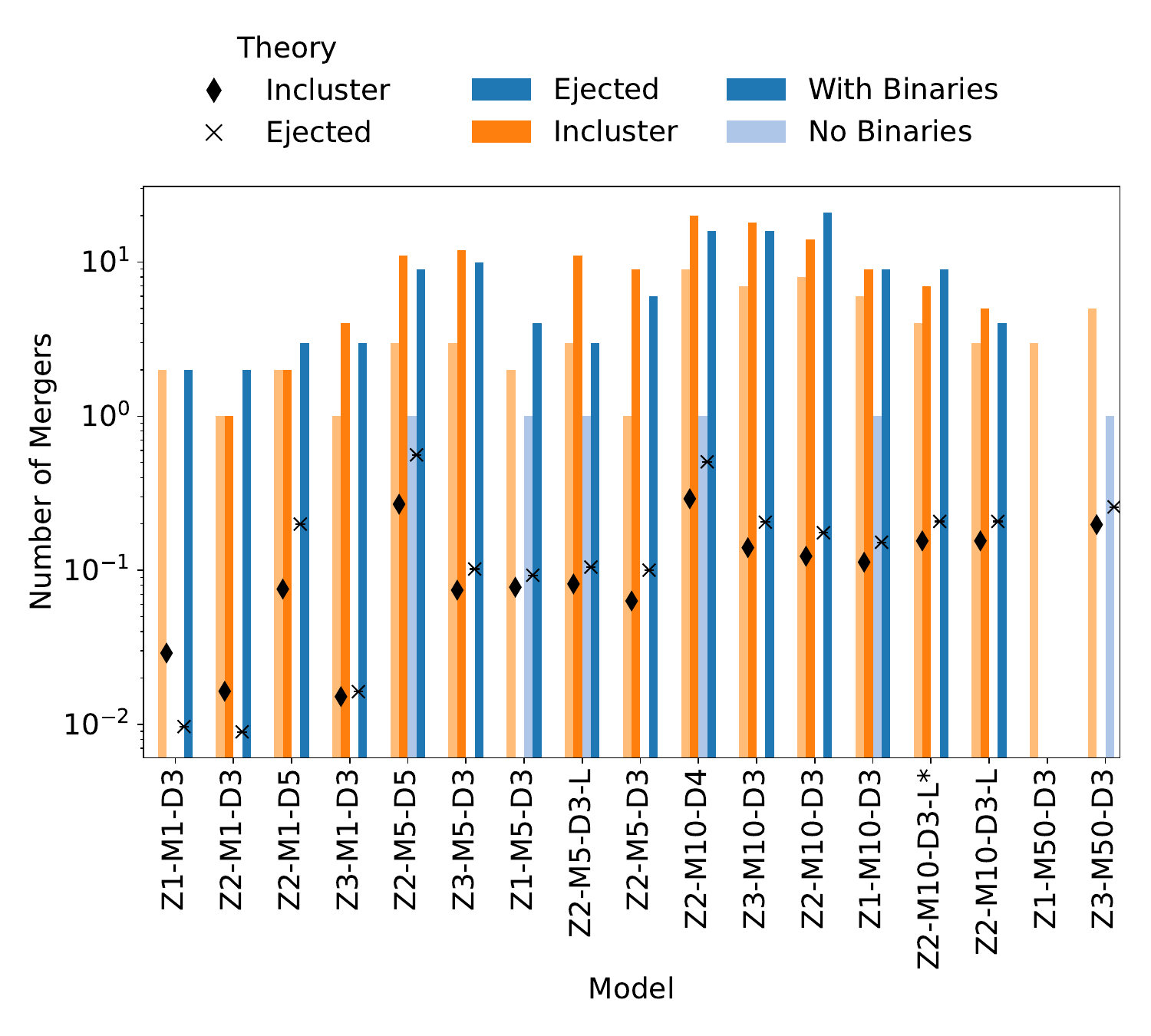}    \includegraphics[width=\columnwidth]{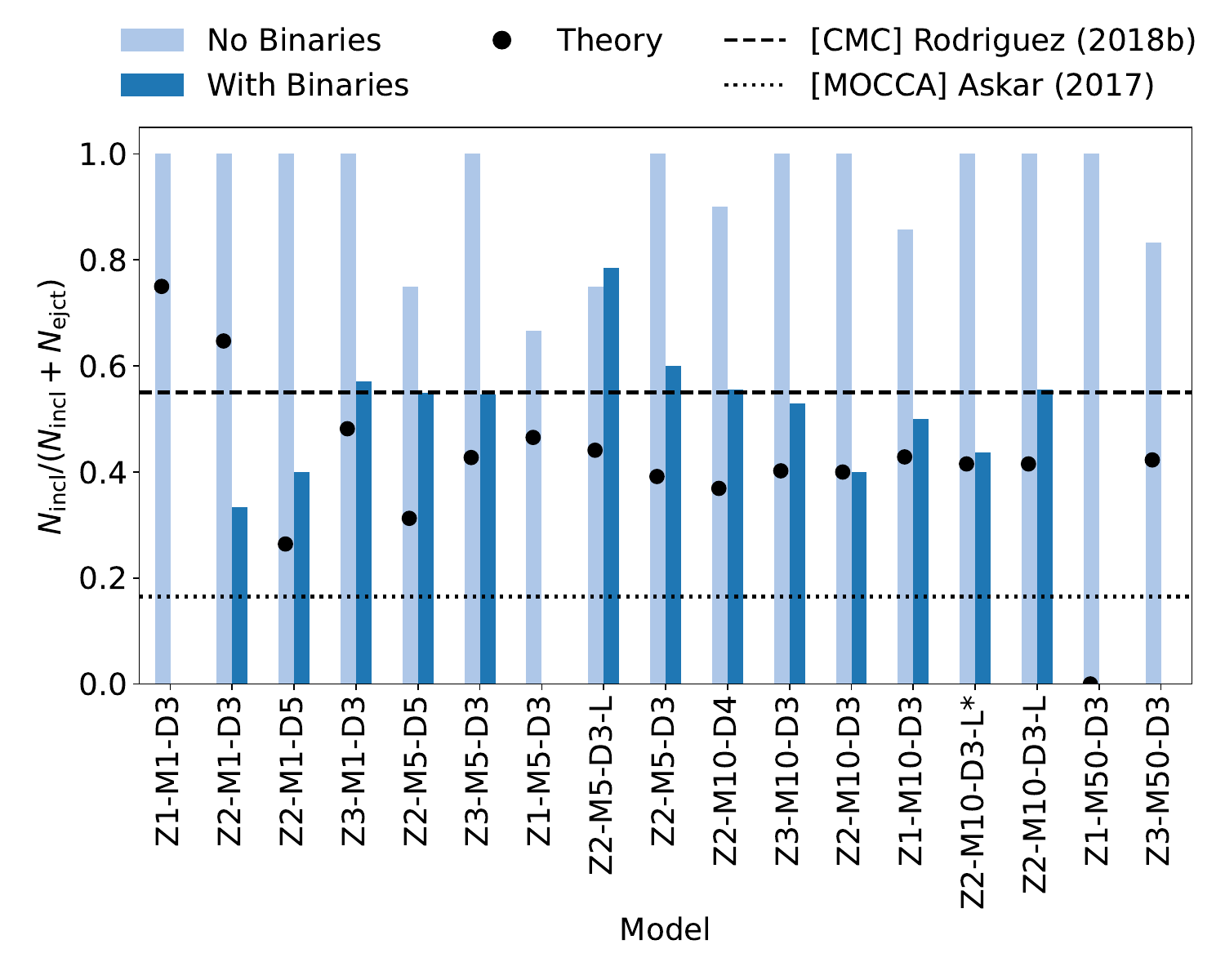}
    \caption{We show the number of ejected and in-cluster BBHs mergers occur in all of our $\tt PeTar$ simulations. Ejected and in-cluster mergers are distinguished by blue and orange bars respectively, whilst the same cluster with and without primordial binaries are shown with darker and lighter shades. We further plot the results from theoretical models for the same initial cluster conditions. On the right panel we show the incluster merger fraction found in two previous studies \citep{rodriguez_post-newtonian_2018-1, askar_mocca-survey_2017} which use monte-carlo cluster codes $\tt CMC$ and $\tt MOCCA$ respectively.
    }
    \label{fig:inclvsEjctAllMods}
\end{figure*}

We show the number of in-cluster and ejected mergers from our simulations in the left panel of Fig.~\ref{fig:inclvsEjctAllMods}. For each model we show the cluster variations with and without primordial binaries and then split up the mergers based on whether they merged in-cluster or were ejected before merger. We also show the average number of in-cluster and ejected mergers from the $\tt cBHBd$ models for each cluster respectively.
In the $\tt PeTar$ models with a primordial binary population, there is a similar number of ejected and in-cluster mergers.
As explained above, most of these binaries are not ejected by dynamical encounters, but by a SN kick during BH formation.
Clusters without primordial binaries exhibit almost no ejected mergers across all simulations. This shows that dynamically formed BBHs tend to merge within the cluster rather than being ejected. The low number of mergers among the ejected population aligns with the results from the $\tt cBHBd$ models, which predict only one merger out of approximately 10 cluster simulations. In line with this, we observe only six mergers, among the 17 cluster models that start without primordial binaries.
In contrast, the predicted number of in-cluster mergers is much higher in the $N$-body simulations compared to the $\tt cBHBd$ models. In total, we find 63 in-cluster mergers, whereas the theoretical expectation is that we should find only  $\mathcal{O}(1)$ merger.

In Fig.~\ref{fig:inclvsEjctAllMods}  we plot the in-cluster fraction ($N_{\mathrm{incl}}/N_{\mathrm{tot}}$) for each simulation. 
As we should expect based on Section~\ref{theory}, the $\tt cBHBd$ models give a $\sim 40\%$ in-cluster fraction.
We note that the lower cluster mass models show results quite different from 40\%, and that this is due to  small number statistics due to the low number of mergers from these clusters. In the $N$-body models without primordial binaries,
essentially all mergers occur inside the clusters.
We conclude that the theory is in disagreement with the $N$-body  model results and that
this disagreement is due to the much larger number of in-cluster mergers produced in the $N$-body models than expected.

It is important to note the $\tt cBHBd$ is a theory based approach that makes specific assumptions about the state of the cluster and how the evolution of the cluster is linked to the formation and evolution of BBHs. In particular, it assumes that once a BBH is formed it only ever experiences strong binary-single interactions which either harden or disrupt the binary. Thus it does not account for higher multiple interactions such as binary-binary interactions, nor does it consider the formation of higher multiplicity systems, i.e., triples or quadruples. 
We showed in a previous section that roughly half of the dynamically formed BBH mergers occurred as the inner binary of a stable BH triple system, and also found quadruple systems in our models. When such higher multiplicity systems are present they  dominate the dynamical interactions due to their large cross section for gravitational encounters.
Moreover, the effect of relatively soft interactions
with closest approach $r_{\rm p}>2a$ are neglected in $\tt cBHBd$ \citep{forastier_binary-single_2024}. As mentioned above, Monte Carlo codes make similar assumptions and also find results that are consistent with $\tt cBHBd$ \citep{fregeau_monte_2007}.
These approximations might at least partly explain the discrepancy between the theory and the full $N$-body simulations.

\begin{figure}
    \centering
    \includegraphics[width=\columnwidth]{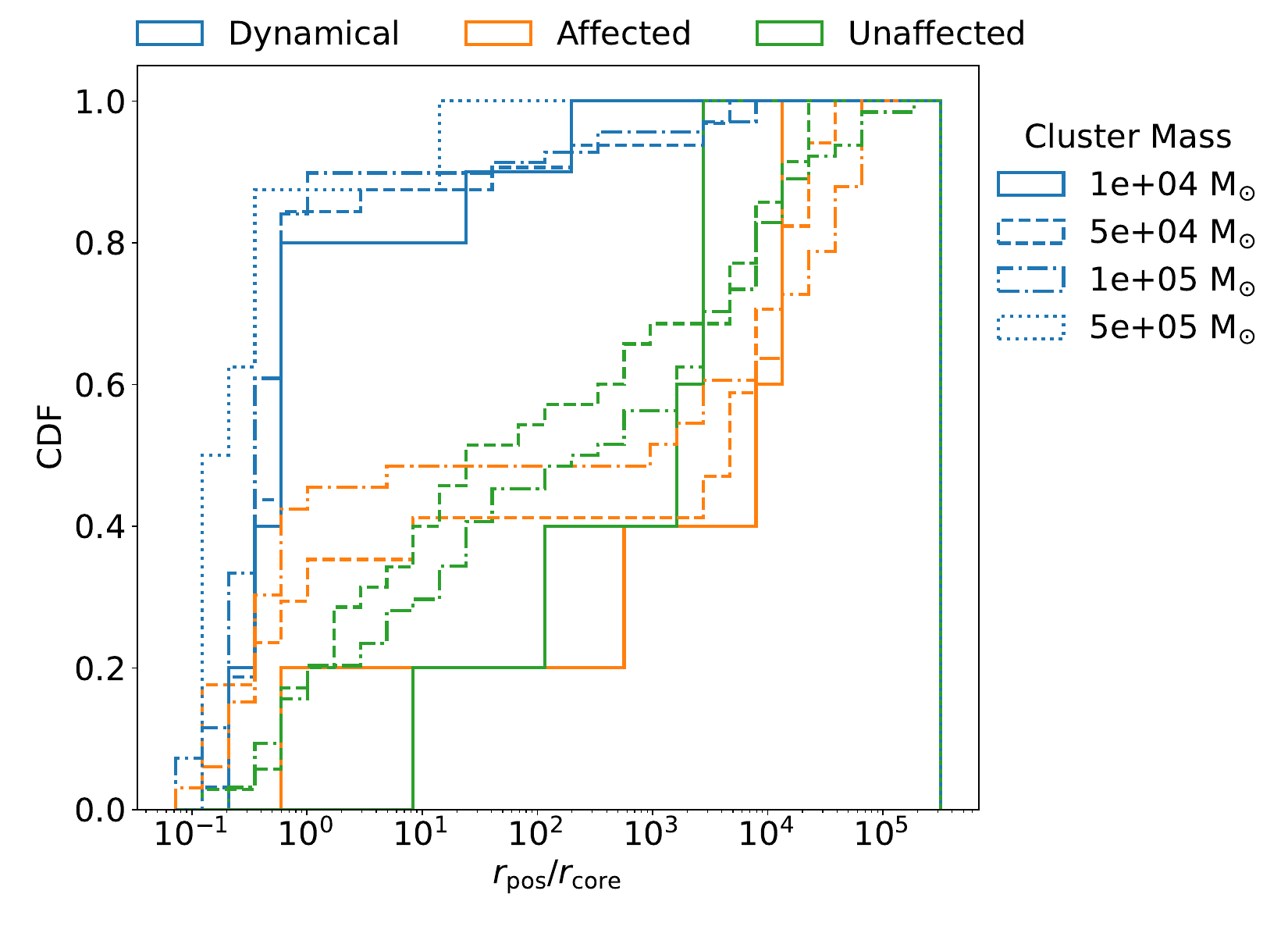}
    \includegraphics[width=\columnwidth]{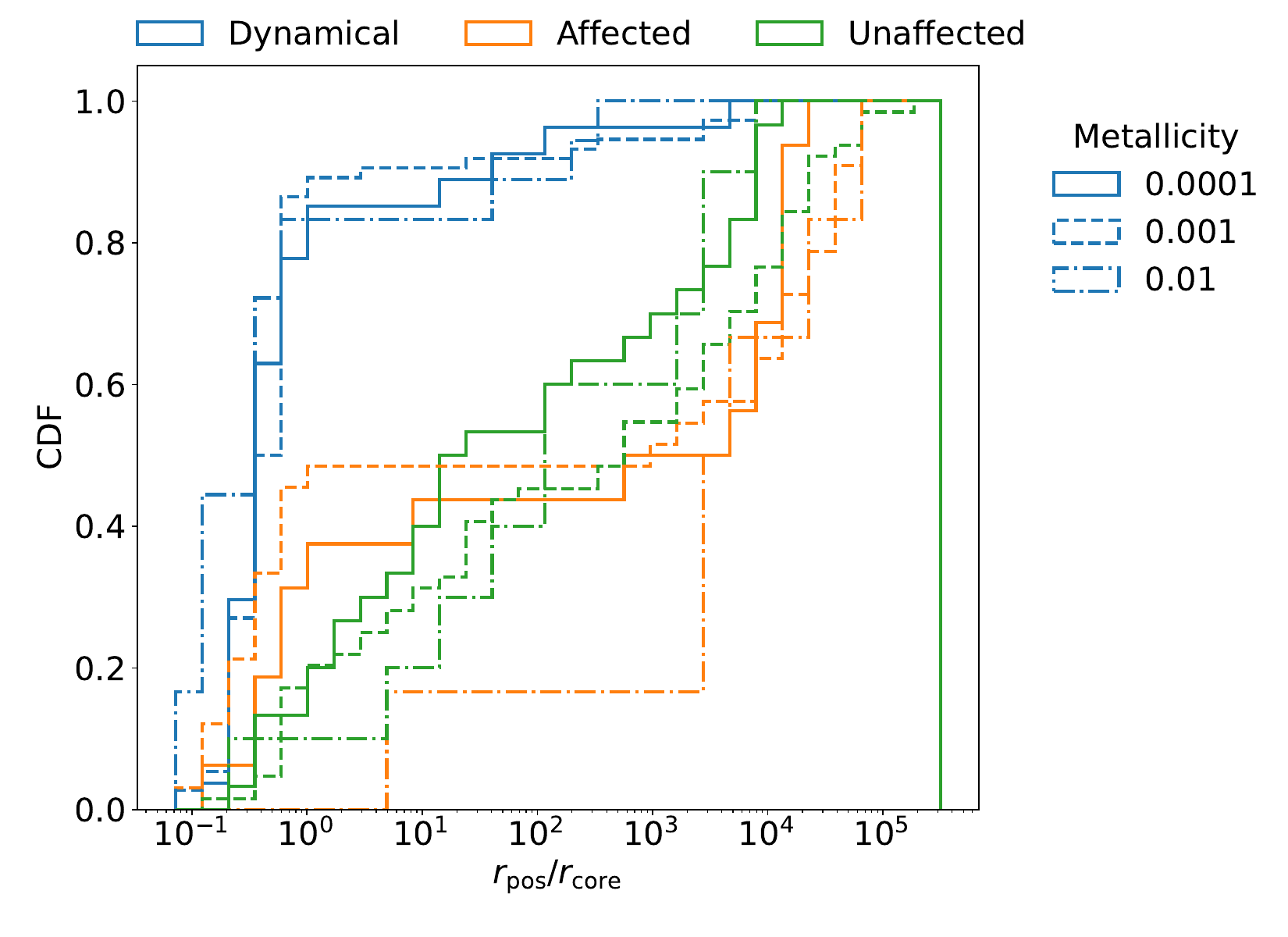}
    \caption{We show the cumulative distribution of the radial position for the affected, unaffected and dynamical BBH mergers in our simulations. The upper panel shows the distribution split by initial cluster mass, whilst the lower panel shows the distribution split by cluster metallicity.}
    \label{fig:radPos}
\end{figure}

To further investigate the in-cluster and ejected merging populations, we look at their radial position at the moment of  merger normalised to the cluster core radius at that time. We split the mergers into the dynamically formed BBHs, the affected primordial BBHs and the unaffected primordial BBHs. Fig.~\ref{fig:radPos} shows the cumulative distribution (CDF) of the radial distance for these mergers, where we have further split the distributions depending on the cluster properties. In the upper panel, we show the CDF for different initial cluster masses and  in the lower panel we are comparing with cluster metallicity. In both plots we find that the cluster properties have little effect on the radial distribution of the mergers.

Focusing on the dynamically formed BBHs,  $84\%$ of the mergers in all simulations occurred within the cluster core, as opposed to $40\%$ for the affected BBHs and $19\%$ for the unaffected BBHs. This supports the idea that these dynamically formed BBHs are forming and merging in the most dynamically active region of the cluster, likely undergoing many  encounters that involve higher multiplicity systems.

\begin{figure*}
    \centering
    \includegraphics[width=\columnwidth]{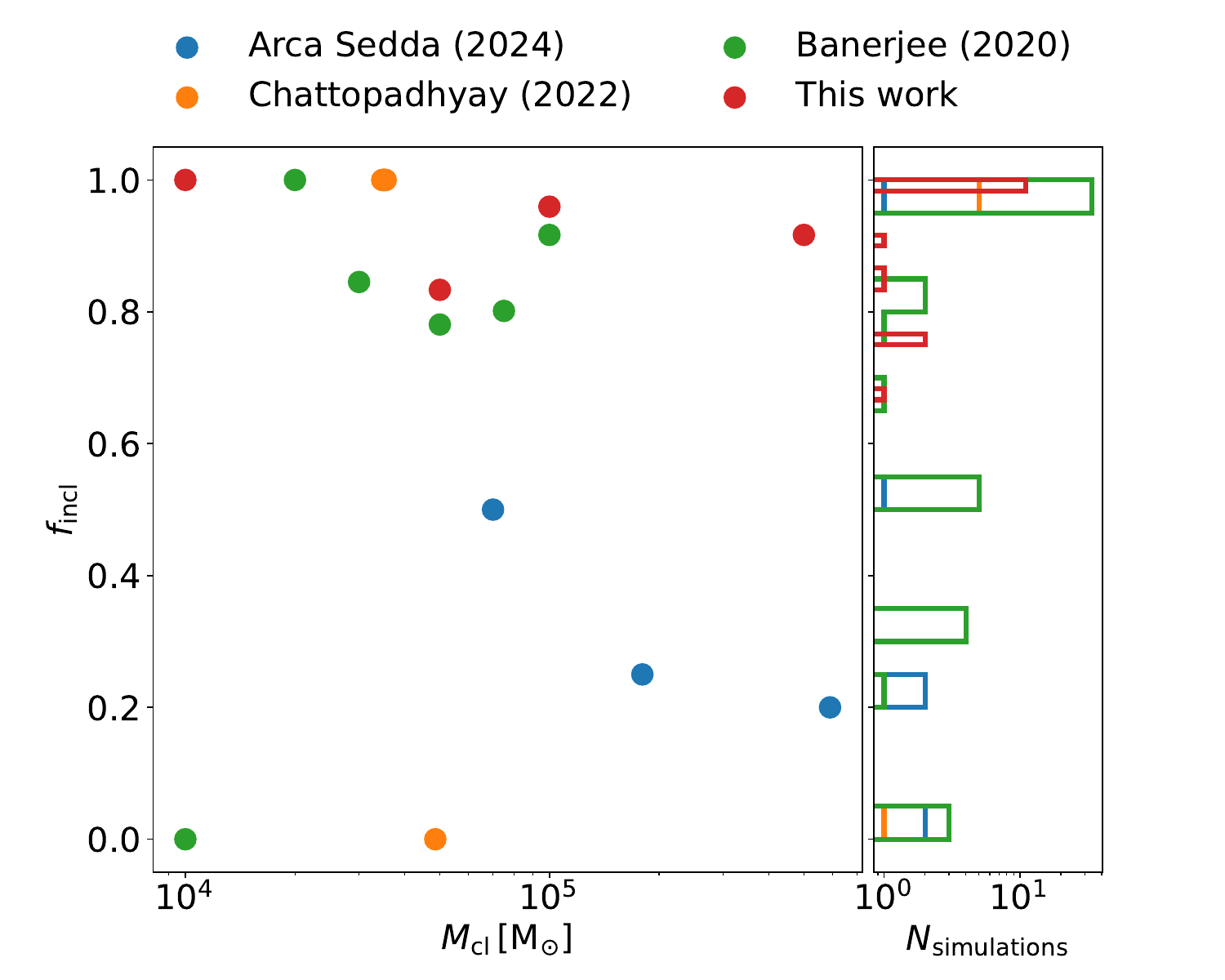}
    \includegraphics[width=\columnwidth]{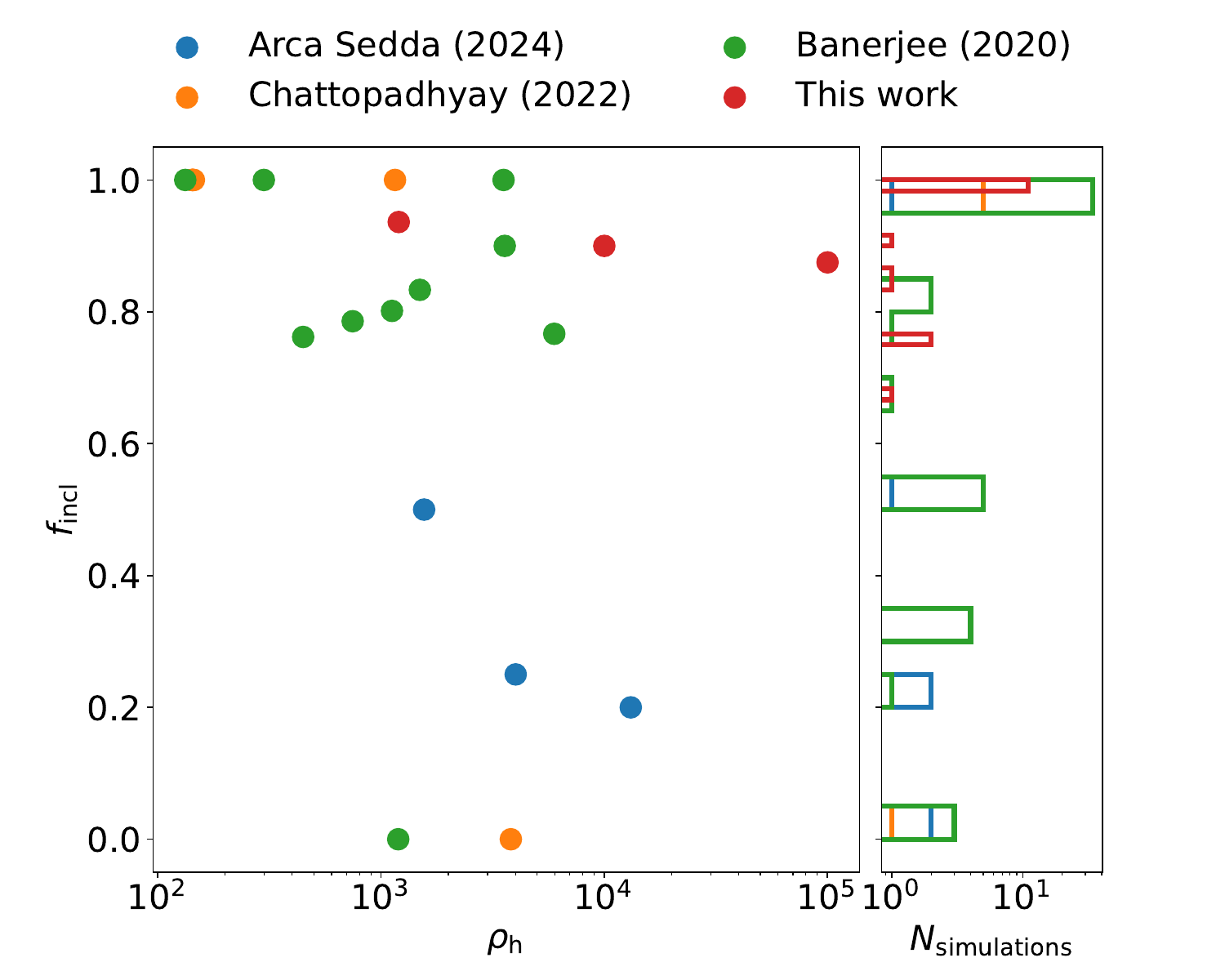}
    \caption{We show the in-cluster fraction of mergers against initial cluster mass (upper panel) and initial density at half mass radius (lower panel). In each plot we include only the clusters with lower primordial binary fraction ($\leq0.1$) and have averaged over all other cluster parameters. We show our results (red) compared to previous works utilising other $N$-body codes \citet{arca2024dragon, chattopadhyay_dynamical_2022, banerjee_stellar-mass_2020}. To the right of each scatter plot we show a histogram of the incluster fraction from every simulation in each of the studies.}
    \label{fig:comparePlot}
\end{figure*}

It is important to put our results into the context of previous star cluster studies. In Fig.~\ref{fig:comparePlot} we plot 
 the in-cluster merger fraction against the initial cluster mass and the  initial cluster half-mass density alongside several previous studies. All of the studies we compare against utilise a direct $N$-body code. \citet{arca2024dragon} use $\tt NBODY6++$ whilst both \citet{chattopadhyay_dynamical_2022} and \citet{banerjee_bse_2020} use $\tt NBODY7$. These codes  include post-Newtonian terms which allows for a self-consistent treatment of general relativistic effects.
Fig.~\ref{fig:comparePlot}
show that our simulations add to the suite of existing work, expanding and filling in more of the parameter space towards the highest mass and density values. 
Since we are interested in understanding the role of dynamics in BBH formation,  we now consider models without primordial binaries, as any BBH in these models must have a dynamical origin.

Many previous studies explore non-zero primordial binary fraction amongst massive stars. Hence, we opt to compare only the in-cluster merger fraction for our clusters with no initial binaries against cluster with $f_{\mathrm{bin}}<10\%$ amongst massive stars from the previous studies. 
We find that our results 
are broadly consistent with most previous studies across both cluster mass and density.  The only 
exception are the simulations by \citet{arca2024dragon}. These authors find that the fraction of in-cluster mergers is $\lesssim 50\%$ in their models. We are unsure about the cause for this difference, but note that all other published models we considered find in-cluster merger fraction that are much higher 
than found by \citet{arca2024dragon} and that are consistent with our results. For example, \citet{chattopadhyay_dynamical_2022} found (3) ejected mergers  in only one of their 11 cluster models. All mergers in their other models  occurred 
inside the cluster.
A key point we find is that in our simulations the in-cluster fraction for the dynamically formed BBHs has no dependence on the initial cluster mass or density. The work  of \citet{banerjee_bse_2020} and \citet{chattopadhyay_dynamical_2022} also do not show any clear dependency,
whilst \citet{arca2024dragon}  find that the number of in-cluster mergers decreases with increasing cluster mass and density.

\section{Formation of high-mass black holes}\label{sec:Massive}
\begin{figure}
    \centering
    \includegraphics[width=\columnwidth]{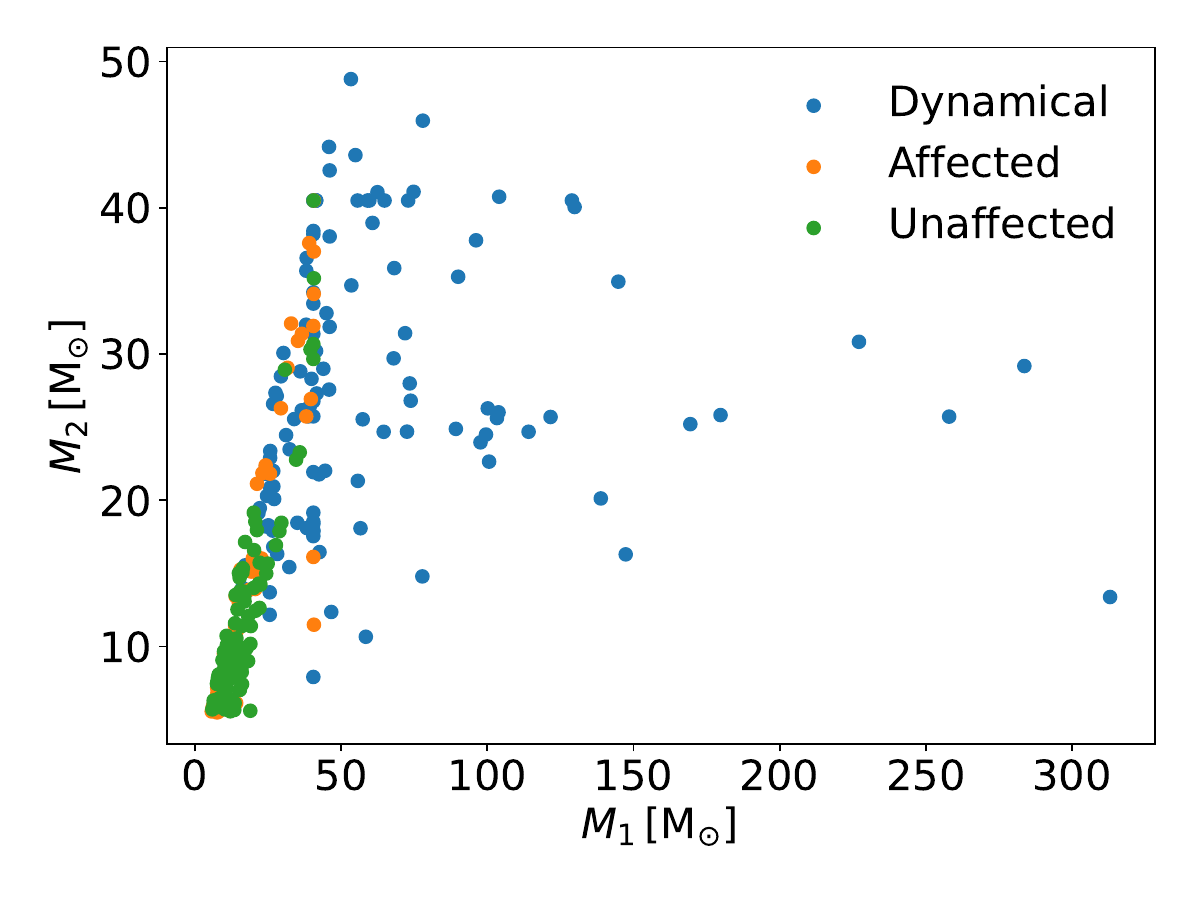}
    \caption{Comparing component masses for all of our BBH mergers, split between the three populations, dynamical binaries, affected primordial binaries and unaffected primordial binaries. Here we have taken the cut-off fractional change defining affected binaries as 50\%.}
    \label{fig:M1vsM2}
\end{figure}

In Fig.~\ref{fig:M1vsM2} we show the primary $vs$ secondary mass for all of our BBH mergers split between the affected, unaffected and dynamical BBH populations. We find  primary BHs with mass $>100\,\mathrm{M_{\sun}}$. These masses  exceed the assumed maximum BH mass that can be formed through stellar evolution in our models. This limit is imposed by the PPSN prescription used in {\tt BSE}, which here is  at $45\,\mathrm{M_{\sun}}$. The BHs with a mass above this limit must have been formed through consecutive mergers with either other BHs or stars \citep[e.g.,][]{zwart_iii_1999,portegies_zwart_formation_2004,mapelli_massive_2016,dicarlo_merging_2019,gonzalez_intermediate-mass_2021, rizzuto_intermediate_2021}. 
In fact, we find that the most massive BHs formed in our models first grow through 
accreting stars, and in a second stage through  mergers with smaller BHs.
These latter mergers are believed to be a key formation mechanism for intermediate and supermassive BH seed growth in massive clusters \citep{antonini_black_2019, chattopadhyay_double_2023}. However, they  require clusters with large $v_{\mathrm{esc}}$ values \citep{antonini_merging_2016}. This is because the asymmetric emission of GWs during a BBH inspiral/merger induces a recoil kick on the remnant BH to conserve momentum. The strength of the kick depends on the mass ratio and spin alignment of the system but can be as large as $\mathcal{O}(10^{3})\,\kms$ \citep{schnittman_distribution_2007}. These recoil kicks are not accounted for in $\tt PeTar$. Therefore, the chain of mergers would have been most likely interrupted after the first BH-BH merger. We can thus consider our simulations as an optimistic upper estimate of the number of mergers from the dynamical BBH population. 

\begin{figure}
    \centering
    \includegraphics[width=\columnwidth]{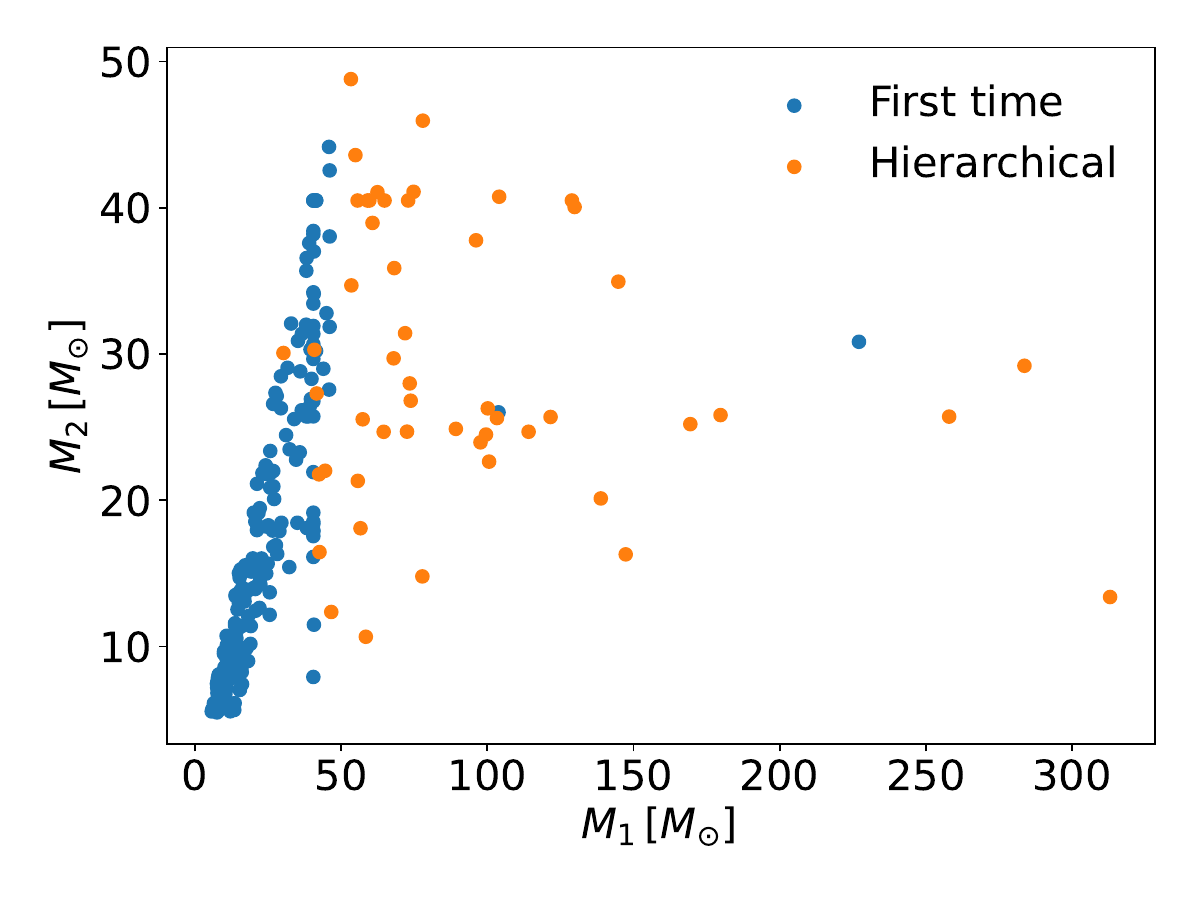}
    \caption{Comparing the component masses for all BBH mergers in our simulations, split by whether this is a first time merger (neither component has been in a BBH merger before) or a hierarchical merger (one or both components are remnants from a previous merger).}
    \label{fig:scatterHier}
\end{figure}

To better understand the formation of these massive BHs we first identify those that have been involved with a least one previous merger. Fig.~\ref{fig:scatterHier} reproduces the scatter between primary and secondary masses, although now we identify two groups based on whether this is a first time merger, or a hierarchical merger where either the primary, secondary or both components have been involved in a previous merger. From Fig.~\ref{fig:scatterHier}, we can see the clear mass limit at $45\,\mathrm{M_{\sun}}$ for the majority of the first time mergers. However, we also see two \textit{first time} mergers which far exceed the PPSN cut-off, one at $\approx100\,\mathrm{M_{\sun}}$ and one at $225\,\mathrm{M_{\sun}}$.

\begin{figure*}
    \centering
    \includegraphics[width=\columnwidth]{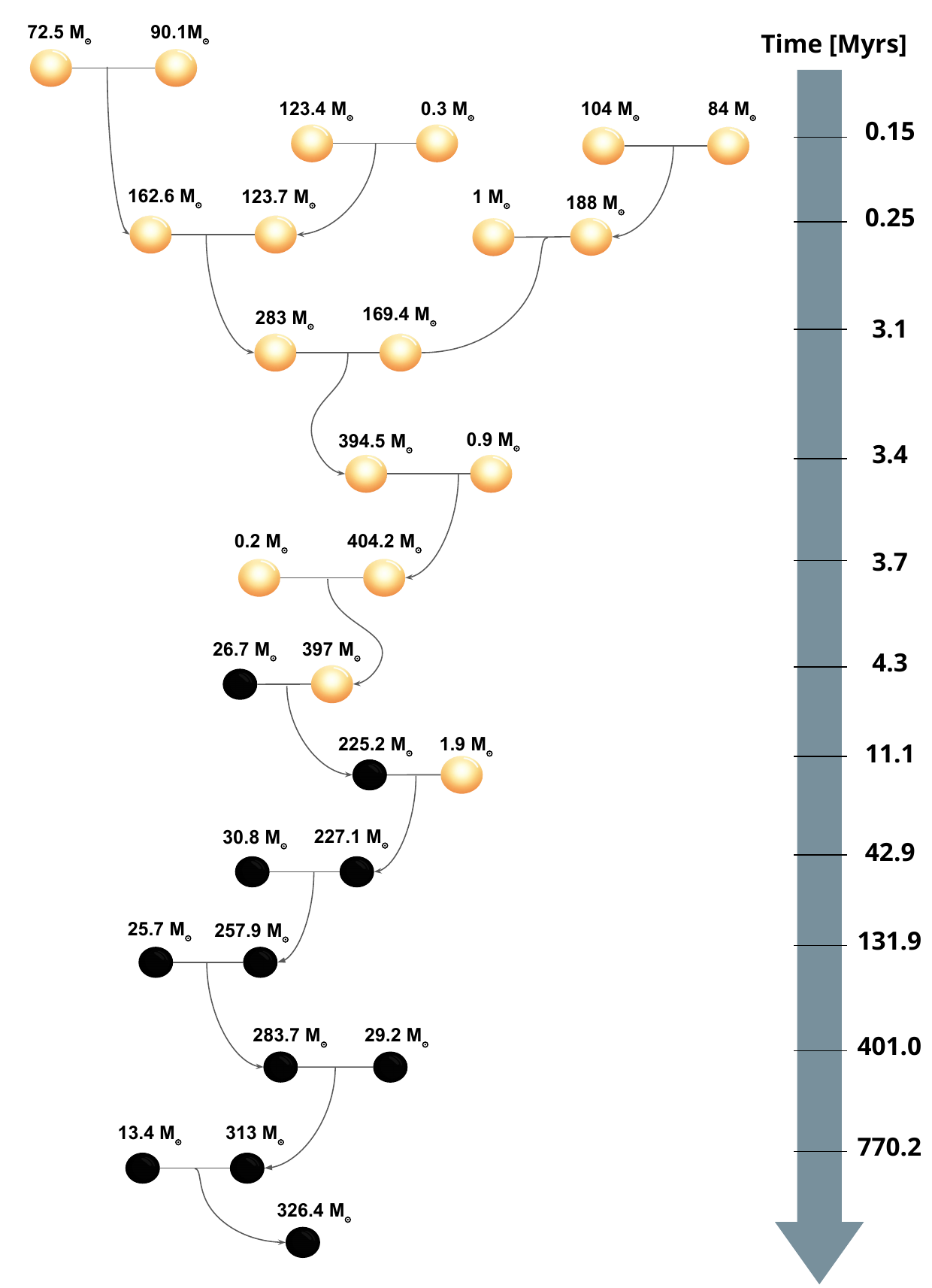}
    \includegraphics[width=\columnwidth]{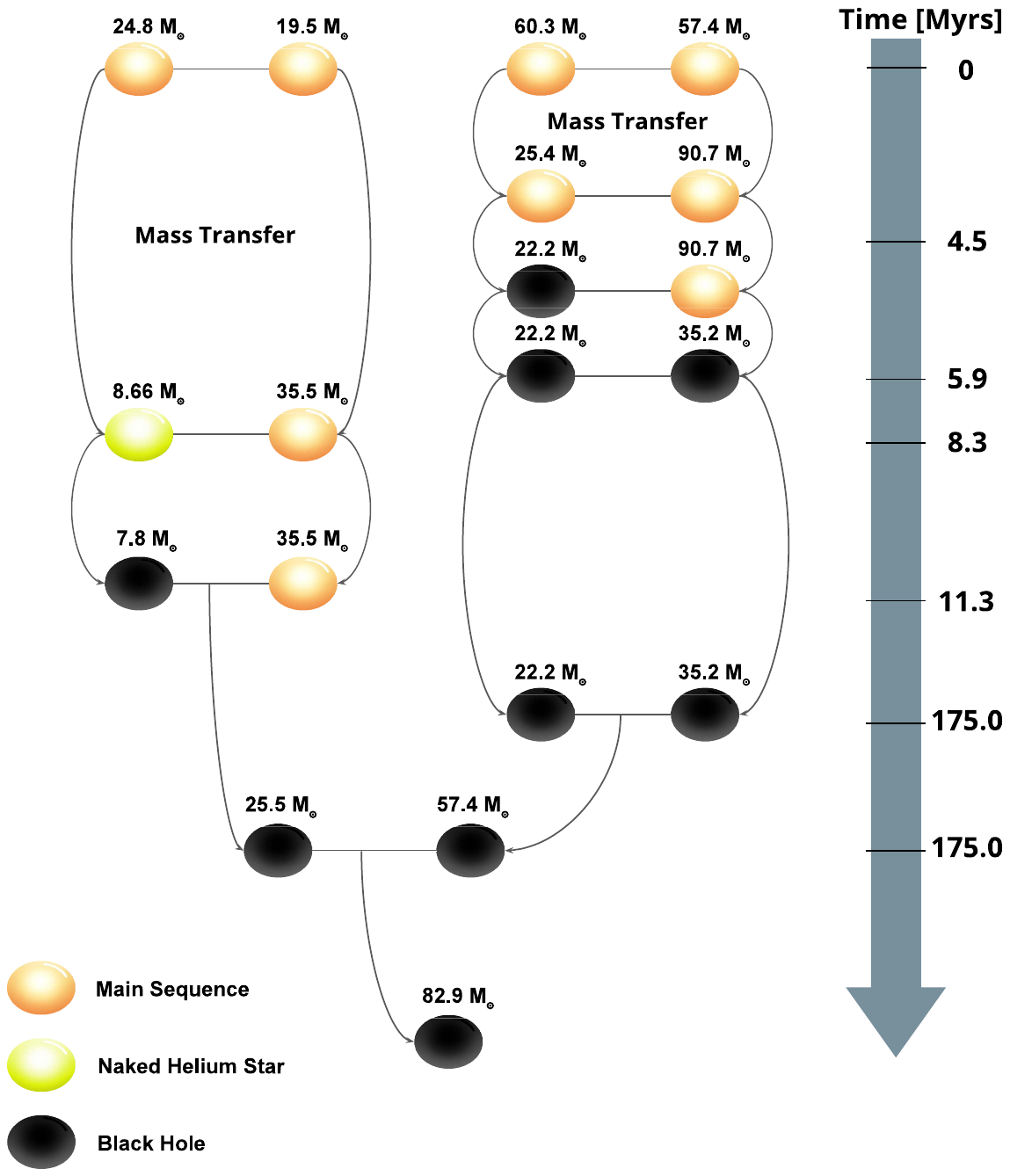}
    \caption{ The left panel show the chain of  mergers leading to the formation of a massive BH  in cluster model Z2-M5-D5. The right panels shows a schematic of the evolution of a primordial binary in model Z3-M5-D3 also leading to the formation of a massive BH. 
    The primordial binary undergoes a mass transfer episode during the stellar phases before merging as a BBH. The remnant BH stays within the cluster and goes on to form a new BBH which then merges within a Hubble time.}
    \label{fig:MassiveBBHSchmatic}
\end{figure*}

We investigate the formation of  the $225\,\mathrm{M_{\sun}}$ BH we track its evolution from the ZAMS of its stellar progenitor up to the end of the simulation. Firstly, we found  this binary in the Z2-M5-D5 cluster model without primordial binaries, thus it is one of the clusters with the highest density. We further found that this massive BH was the result of eight previous stellar mergers which produced a $397\,\mathrm{M_{\sun}}$ star which is then swallowed by a $26.7\,\mathrm{M_{\sun}}$ BH. This results in a $225\,\mathrm{M_{\sun}}$ BH which, following the merger with one final star of mass $2\,\mathrm{M_{\sun}}$, forms a BBH and eventually merges. Notably, we find that the remnant of this BBH merger then goes on to form another merging BBH three more times. We note that a caveat to this evolutionary pathway is the fact that $\tt PeTar$ does not currently model mass loss during stellar collisions. For massive stars with loosely bound envelopes it is likely that stellar collisions remove significant mass from the star, thus restricting mass growth. We show a schematic of these mergers in the left panel of Fig.~\ref{fig:MassiveBBHSchmatic} along with the masses of all the components. We also note that in this case, GW recoil kicks  are unlikely to be  large enough to eject the remnant from the cluster due to the low mass ratio of the BBH \citep{holley-bockelmann_gravitational_2008}. 
Since this merger was found in a cluster with no primordial binaries, we then choose another hierarchical merger from a cluster model containing a primordial binary population to compare the evolutionary pathways. We opt for a hierarchical merger with the $M_{\rm 1}=57.4\,\mathrm{M_{\sun}}$ and $M_{\rm 2}=25.5\,\mathrm{M_{\sun}}$. We find that the secondary BH is the result of the evolution of a primordial binary system, which undergoes some period of mass transfer before one component forms a BH and then quickly merges with its companion star. On the other hand, the primary BH is the remnant of a previous BBH merger where the binary was from the primordial binary population, evolved together, formed a BBH and merged. The schematic from this merger chain is shown in the right panel of Fig.~\ref{fig:MassiveBBHSchmatic}. 

\section{Conclusions}\label{sec:conclusion}
In this work we ran 35  $N$-body simulations of stellar clusters using the $N-$body code $\tt PeTar$. We produced stellar clusters which span a range of initial cluster masses from $10^{4}\,\mathrm{M_{\sun}}$ to $10^{6}\,\mathrm{M_{\sun}}$, initial half-mass density from $1200\,\density$ to $10^{5}\,\density$ and metallicity values $0.0001$, $0.001$ and $0.01$. For each cluster simulation, we ran two variations, one with no primordial binaries, and another with 100\% binary fraction amongst massive ($\geq20\,\mathrm{M_{\sun}}$) stars (see Table~\ref{tab:initCond}). We investigated the population of BBH mergers, identifying the impact of the cluster environment and dynamical interactions on the binary  properties and merger rate.
We compared the results of our simulations to the predictions based on our theoretical understanding of BH dynamics in clusters. Our main conclusions are summarised in what follows:
\begin{itemize}

\item[(1)] in clusters that start with a realistic population of massive binaries, the majority of BBH mergers originate from the primordial binary population rather than being paired by dynamical interactions (see Fig.~\ref{fig:BHNumbers}, Fig.~\ref{fig:RadvsTime},  and Fig.~\ref{fig:affctBins}). 
\item[(2)] This primordial BBH merger population  is composed of two groups. One group is unaffected by the dynamical environment of the cluster either due to being ejected from it early on due to a natal kick, or because they are initially tight enough to merge before any interaction. The other group remains in the cluster for some time and undergoes at least one encounter which changes their orbital properties. We find that    about $20\%$ of all primordial BBH mergers are  significantly \textit{affected} by dynamics, in the sense that their merger timescale changes by at least a factor of 2 due to dynamical interactions  (see Fig.~\ref{fig:affctBins}). The two populations are characterised by statistically different distributions of component masses, delay times, eccentricities and mass-ratios (see Fig. \ref{fig:BBHDists}).
\item[(3)] Due to the subdominant number  of BBH mergers that are formed or affected by dynamical encounters in the cluster,  the overall merger rate  from the  $N$-body models is essentially the same as if the  cluster stars were evolved in isolation (see Fig. \ref{fig:mergeEff}).
\item[(4)] Conclusion (3) depends on both the assumed prescription of common-envelope evolution and metallicity.   If binaries are assumed to merge when the CE is initiated by a star crossing the HG, then the merger rate for $Z=0.01$ is increased due to dynamical interactions by $\simeq 3$ orders of magnitude. Under the same assumption and for $Z=0.0001$, the effect of dynamics on the number of BBH mergers remains negligible (see Fig. \ref{fig:mergeEff}).
\item[(5)] 
Almost all BBH mergers that are formed dynamically  merge while the binary is still inside the parent cluster.  
This is in contrast to the theoretical expectation that  about half of the mergers should occur outside the cluster. We argue that this is  due to  encounters
involving systems with higher hierarchy such as triples and quadruples and/or multi-body interactions
(beyond binary-binary and binary-single encounters) that are neglected in semi-analytical and Monte Carlo codes.
\item[(6)] We did not observe a clear correlation between the number of dynamically formed BBH mergers with cluster mass or density. 
\item[(7)] We searched  for higher multiplicity systems and found that 
 $\simeq 10\%$  of the merging primordial BBHs are the inner binary of a stable triple BH system. In contrast,  dynamically formed BBHs can be found in approximately equal numbers  in binaries and (stable) triples, as well as a  small fraction $<1\%$ in quadruples. 
 \item[(8)] We find several hierarchical BBH mergers with primary masses $>45\,\mathrm{M_{\sun}}$, although these are overproduced in our models due to the lack of a GW recoil prescription in {\tt PeTar}. However, we also find two instances of a first time BBH merger where the primary mass is above the PPSN mass gap. Tracking the history of the most massive case we found the formation path for the primary involved several stellar mergers in succession, producing a massive star ($M=397\,\mathrm{M_{\sun}}$) which is then swallowed by a BH ($M=26.7\,\mathrm{M_{\sun}}$). This results in a very massive BH ($M=225.2\,\mathrm{M_{\sun}}$) which then goes on to form a BBH and merge. This presents a potential mechanism for producing massive BHs well above the PPSN mas limit.
\end{itemize}

The  simulations presented in this work build on the collection of existing cluster simulations, exploring a more extreme region of parameter space than done before. In particular we explored clusters with masses and densities that are comparable to those of present-day globular clusters and with an observationally motivated initial binary fraction. We compared the in-cluster merger fraction from our models with no primordial binaries, against previous studies. We show that our results are broadly consistent with the  work of \citet{banerjee_bse_2020} and \citet{chattopadhyay_dynamical_2022}, while they  differ from \citet{arca2024dragon} who find a much lower in-cluster fraction. It is important to note that when comparing N-Body studies, the choice of initial conditions can significantly impact BBH formation and evolution. For example, in \citet{arca2024dragon}, the mergers of primordial BBHs occur almost entirely outside the cluster, whereas in our work, we found a more even split between ejected and in-cluster mergers. One key difference between these studies lies in the initialization of the simulations, with \citet{arca2024dragon} opting for a lower cluster concentration ($W_{0}=6$). This implies a lower central escape velocity for a given cluster mass and density,  suggesting that binaries are   more susceptible to removal from the cluster through natal kicks, increasing the number of mergers among the ejected population.

Our findings have several significant implications. Firstly, they indicate that dense and massive clusters as the ones considered in this work might account for only a small portion of the overall BBH merger rate in the Universe. This is because  most stars do not form within such  dense  clusters, and the merger rate in our models with a primordial binary population is not substantially increased by the binary's presence in a dense cluster.
We stress, however, that this conclusion is based on the results of massive binary evolution calculations that remain quite uncertain. Moreover, we showed that dynamically formed  BBH mergers have  larger masses and eccentricities than those formed in isolation, making them a distinct and possibly identifiable population of mergers \citep[see also][]{rodriguez_redshift_2018,di_carlo_binary_2020,torniamenti_dynamics_2022,belczynski_black_2022}.
 The lack of  a clear correlation between the 
 number of dynamically formed BBH mergers with cluster properties is also interesting, and deserves further investigation which will require simulations extending the parameter space  to even higher masses and densities.

Finally, our results cast doubts on conclusions derived from simplified models of cluster dynamics where binary-single and binary-binary encounters are assumed to be the main form of  interactions leading to BBH mergers. This is especially relevant to GW detections, as the fraction of eccentric BBH mergers originating from clusters is expected to scale with the number of in-cluster mergers. It is typically stated that $\sim 5\%$ of mergers from clusters will have a residual eccentricity at the moment they enter the $\gtrsim 10$ Hz window of current detectors \citep{samsing_eccentric_2018}. The large number of in-cluster mergers found in our models is likely to imply a much higher number of eccentric mergers. Wile the reason for the discrepancy  is unclear at the moment, we plan to carefully investigate this in future work.

\section*{Acknowledgements}
We thank Long Wang for useful discussions and for assisting with setting up and running the $N$-body simulations. We thank Debatri Chattopadhyay and Mark Gieles for useful suggestions and comments.
Simulations in this paper utilised the high-performance $N$-body code $\tt PeTar$ (version 1047-\_290) which is free available at \url{https://github.com/lwang-astro/PeTar}. We acknowledge the support of the Supercomputing Wales project, which is part-funded by the European Regional Development Fund (ERDF) via Welsh Government. JB is supported by the STFC grant ST/T50600X/1 and FA is supported by the UK’s Science and Technology Facilities Council grant ST/V005618/1.

\section*{Data Availability}
The data used for this work will be freely shared upon reasonable request to the author.



\bibliographystyle{mnras}
\bibliography{Refs} 








\bsp	
\label{lastpage}
\end{document}